\documentclass[aps,prl,superscriptaddress,twocolumn,nobibnotes]{revtex4-2}

\usepackage{graphicx}
\usepackage{epstopdf}
\usepackage{comment}
\usepackage{textcomp}
\usepackage{gensymb}
\usepackage{amsmath,amssymb}                
\usepackage{color,colordvi}
\usepackage{hyperref}
\usepackage{xcolor}
\usepackage{graphicx}
\usepackage{subcaption}
\usepackage{amssymb}
\usepackage{float}

\begin{document}
\title{Star-like microgels vs star polymers: similarities and differences}
\author{Tommaso Papetti}
\email[Corresponding author:]{tommaso.papetti@uniroma1.it}
\affiliation{Department of Physics, Sapienza University of Rome, Piazzale Aldo Moro 2, 00185, Roma, Italy}
\affiliation{CNR Institute of Complex Systems, Uos Sapienza, Piazzale Aldo Moro 2, 00185, Roma, Italy}
\author{Elisa Ballin}
\affiliation{Department of Physics, Sapienza University of Rome, Piazzale Aldo Moro 2, 00185, Roma, Italy}
\affiliation{CNR Institute of Complex Systems, Uos Sapienza, Piazzale Aldo Moro 2, 00185, Roma, Italy}
\author{Francesco Brasili}
\affiliation{CNR Institute of Complex Systems, Uos Sapienza, Piazzale Aldo Moro 2, 00185, Roma, Italy}
\affiliation{Department of Physics, Sapienza University of Rome, Piazzale Aldo Moro 2, 00185, Roma, Italy}
\author{Emanuela Zaccarelli}
\email[Corresponding author:]{emanuela.zaccarelli@cnr.it}
\affiliation{CNR Institute of Complex Systems, Uos Sapienza, Piazzale Aldo Moro 2, 00185, Roma, Italy}
\affiliation{Department of Physics, Sapienza University of Rome, Piazzale Aldo Moro 2, 00185, Roma, Italy}

\begin{abstract}
Star-like microgels have recently emerged as a promising class of thermoresponsive soft colloids, that have an internal architecture similar to that of star polymers. Here, we perform extensive monomer-resolved simulations to theoretically establish this analogy. First, we characterize the effective potential between star-like microgels, finding that it is Gaussian for an extended range of distances, in stark contrast to the Hertzian-like one of standard microgels, but almost identical to that of star polymers with a core partially covered by chains. Next, we investigate the ratio between gyration and hydrodynamic radii across the volume-phase transition, showing qualitative agreement with both star polymers and experimental data. Finally, we estimate the bulk modulus, finding star-like microgels significantly softer than standard microgels and comparable to star polymers. The present work thus demonstrates that star-like microgels behave as ultrasoft particles, akin to star polymers, paving the way for their exploration at high concentrations.
\end{abstract}
\maketitle

\section{Introduction}
One of the key advantages to work with colloidal systems is their tunable \textit{softness}, that can be modified by varying the internal architecture of the particles. A pioneering review article by Vlassopoulos and Cloitre~\cite{vlassopoulos2014tunable} has proposed to define softness as the ratio of the elastic free energy over the Boltzmann factor, $s = F_{el}/k_bT$. This allows to classify colloidal particles from very soft, i.e., polymer chains with $s\sim O(1)$, to completely hard (hard spheres), with $s \rightarrow \infty$. In between these extreme cases, there exists a variety of particles, often bridging together polymeric and colloidal nature, giving rise to a range of intermediate softness. Among these, we focus in this work on  star polymers ($s\sim 10^3$) and microgels ($s\sim 10^4$). 

Star polymers are colloidal particles made of long polymer chains, generally called arms, that are chemically attached to a common center, the core, whose dimension is much smaller than the arm length. Their behavior is essentially governed by the arm number $f$ and has been extensively studied in the past both theoretically~\cite{likos2001effective,likos2006soft} and experimentally~\cite{vlassopoulos2004colloidal,erwin2010dynamics}. 
The classical theoretical description of star polymers is based on the well-established Daoud-Cotton blob model, where correlated blobs of increasing size depart from the central core.  
With this description, it was theoretically shown and numerically validated that the effective interaction potential between two stars, using as a natural variable the distance between the cores, increases logarithmically at short distances~\cite{witten1986colloid,likos1998star}. Thanks to this very slow divergence at full contact, the potential between star polymers is often referred to as ultrasoft. At longer distances, there is not an analytic description of the potential, which was postulated to bend over 
a Yukawa form for $f >10$~\cite{likos1998star}. At small enough $f$, a Gaussian description seems to be more accurate~\cite{jusufi2001effective}.
These predictions were found to be in agreement with experiments~\cite{likos1998star} and computer simulations~\cite{jusufi2001effective} for moderate values of arm numbers. However, star polymers can also be considered as deformable colloids and is possible to use the distance between their centers of mass as the variable to calculate their effective potential, similarly to what is commonly done with other macromolecules. In this case, it was shown by theoretical calculations and simulations on a lattice that the effective potential is not logarithmic, but rather adopts a Gaussian form~\cite{hsu2004effective} which seems to hold in the whole range of relative distances.

A drawback acting against a wide use of star polymers in the experimental community is the rather involved synthetic process, which led in recent years to the search of alternative model systems with comparable properties. One notable example is given by asymmetric block copolymer micelles, assembling into star-like objects~\cite{laurati2005starlike}, whose phase behavior and interactions were found to be rather similar to that of ideal stars~\cite{gupta2015dynamic,gupta2015validity}.

On the other hand, microgels are crosslinked polymer networks, whose softness is normally controlled by the amount of crosslinkers $c$. Usually, they are synthesized by making use of thermoresponsive polymers, such as Poly-N-isopropylacrylamide (PNIPAM), providing the microgels with the ability to respond to temperature. In particular, PNIPAM microgels are well-known to undergo a reversible Volume Phase Transition (VPT) at a temperature $\sim 32 \,^\circ \mathrm{C}$, which makes them very suitable for a variety of applications~\cite{karg2019nanogels}. The standard preparation method is a simple precipitation polymerization, routinely employed by several groups around the world, which has favoured their spreading in the community as a favourite model system for understanding the behavior of soft and deformable objects~\cite{yunker2014physics,philippe2018glass}. Traditionally, bisacrylamide (BIS) is the most used crosslinking agent. Since it reacts a bit faster than NIPAM, it yields the microgels with a characteristic internal structure, that is well-described by a fuzzy sphere model~\cite{stieger2004small}. This amounts to a denser core, rich in crosslinkers, and to a soft corona, poor of crosslinkers.

However, a recent work has put forward a new class of soft particles, which has the potential to bridge the gap between star polymers and microgels.
These particles, named star-like microgels~\cite{ballin2025star}, are still thermoresponsive microgels based on PNIPAM, synthesized in an identical way to standard ones but with the only difference that the BIS crosslinker is substituted by another crosslinking agent: ethylene glycol dimethacrylate (EGDMA)~\cite{kratz2002volume}. The latter has a much higher reactivity than BIS, accumulating within a narrow central region of the microgels, analogous to the core of the star, and leaving the rest of NIPAM assembling into chains of various lengths, i.e., the arms. Ref.~\cite{ballin2025star} reported a joint experimental and numerical investigation of such microgels, aimed to characterize their internal structure under dilute conditions as a function of temperature across the VPT. The measured Small Angle X-Ray Scattering (SAXS) form factors of the microgels were described with a hybrid model, combining star and microgel features, the so-called core-fuzzy sphere model,  for two representative values of the EGDMA crosslinker concentration. For $c=1\%$, the microgels were found to retain star-like character, while for $c=10\%$, the core becomes much bigger and a simple star-like structure does not hold any longer. The data were then compared to monomer-resolved simulations of ideal star polymers, confirming a good agreement with ideal stars with $f=80$ for $c=1\%$. To make a step forward, the \textit{in silico} synthesis of realistic monomer-resolved microgels, put forward in Refs.~\cite{gnan2017silico,ninarello2019modeling}, was also extended to describe the internal structure of PNIPAM-EGDMA microgels, finding remarkable agreement with experiments.

However, a fundamental question remains open, regarding the true similarity of star-like microgels with low EGDMA content to ideal star polymers. To answer this, in this work we exploit our \textit{in silico}  model to evaluate the effective interactions $V_{eff}$ between star-like microgels with $c=1\%$ and compare them with those of star polymers. Given that in the literature, numerical calculations of $V_{eff}$ are not available for stars with large arm numbers, in the range needed to make a comparison with the present star-like microgels, we also provide new calculations of the star effective potentials at different arm numbers. On the other hand, effective interactions between standard microgels are usually reported to be Hertzian-like~\cite{rovigatti2019connecting,del2024numerical}, as typical of elastic spheres.
We thus aim to shed light on the interactions between star-like microgels, as also compared to standard microgels and classify them among the variety of soft colloids. 

Interestingly, we find that the calculated $V_{eff}$ for star-like microgels is Gaussian, confirming their star-like character, but the resulting repulsion is significantly lower than for stars. We attribute this to the lower coverage of the core with respect to ideal stars achieved in PNIPAM-EGDMA microgels and thus compare our findings to a partially covered star, whose effective interactions are found almost identical to those of star-like microgels. In addition, we provide an estimate of the bulk modulus, finding that the star-like microgels are again similarly soft to star polymers, and much softer than regular microgels.
In order to further validate the numerical model, we also report an estimate of the ratio between the gyration radius $R_g$ and the hydrodynamic radius $R_H$ as a function of temperature for all the simulated systems  and compare them with new experimental measurements for star-like microgels, performed by the combination of Static and Dynamic light scattering. 

The manuscript is organized as follows. After describing our numerical and experimental methods, we report the calculated effective potentials for star polymers, for star-like microgels, for regular microgels and for partially covered stars, comparing them to the known analytical and phenomenological forms and among themselves. We then also provide numerical estimates for $R_g/R_H$  for all these systems and compare them to literature for star polymers and to new measurements performed for star-like microgels. For the latter, we report the full temperature dependence in comparison to simulations, where this is varied by introducing a solvophobic attraction between the monomers. %
We then discuss our main findings in the context of existing literature, finding a strong analogy between star-like microgels and star polymers having a lower core surface coverage than ideal ones, as also confirmed by the analysis of the bulk modulus of the particles. Incidentally, we also show that this reduced coverage of the core surface does not qualitatively affect the shape of the effective interactions and the internal structure of the particles with respect to regular stars.

The present work thus theoretically establishes star-like microgels as a promising model system for investigating the role of softness in the behavior of dense colloidal systems, akin to star polymers and also able to respond to temperature.

\section{Numerical Methods}

\label{sec:num_methods}
\subsection{Star Polymer Model}
We construct a star polymer by placing a single central monomer with a finite radius $R_c$ and decorating it with $f$ arms, each made of $N_f$ monomers of diameter $\sigma$ and unit mass $m$. We vary $f$ in the range from 6 to 80, the latter value being the estimate of the arm number for star-like microgels established in Ref.~\cite{ballin2025star}. For the latter case, we consider  $N_f=200$, but we also report selected results for $N_f=50$. 
The effective coverage fraction, defined as $\gamma = f/16R_c^2$, is set to 1~\cite{likos2001effective}, hence $R_c\simeq 2.2 \sigma$. 
However, to compare with star-like microgels, we also perform simulations of non-ideal stars, where we vary the value of $\gamma$. That is accomplished either by covering the core (with the same size as above) with a smaller number of arms, or by fixing the latter and changing the core radius in the range $1.7\,\sigma \lesssim R_c \lesssim 4 \,\sigma$. In all cases we start with a configuration where arms are fully stretched, that is later equilibrated with the monomer interactions described below.

\subsection{Microgel model}
The assembly of the microgels is performed as in previous works~\cite{gnan2017silico,ninarello2019modeling}. This is by now a well-established procedure, that exploits the assembly of bivalent patchy particles (the monomers) and tetravalent patchy particles (the crosslinkers), all with diameter $\sigma$ in a spherical volume, with number density $\rho\sim 0.08$, fixed by the favourable comparison with experiments~\cite{ninarello2019modeling}. During the assembly, the crosslinkers experience an additional force, which phenomenologically represents their higher reactivity, which accumulates them preferentially in the center of the microgels. This force is found to be always the same for PNIPAM-BIS microgels with any $c$ and was generalized in Ref.~\cite{ballin2025star} to the case of PNIPAM-EGDMA microgels. In addition, EGDMA monomers can form bonds between themselves, while for BIS ones this possibility is usually neglected. 

Using these protocols, here we assemble star-like microgels starting with $N_m=21000$ particles, of which 1\% is the molar amount of crosslinkers. For comparison, we also prepare \textit{standard} (also sometimes called \textit{regular}) microgels, having the same crosslinker concentration and a similar $N_m$. These microgels have a standard fuzzy sphere architecture as those studied in our previous works~\cite{ninarello2019modeling,marin2025predicting}. In all cases, when most bonds ($>99.5\%$ of the total) are formed, we select the largest cluster as the obtained microgel networks, whose topology is then fixed. The resulting microgels have $N_m\sim 20000$ for regular microgels and $N_m\sim 16000$ for star-like microgels.

\subsection{Interaction Potentials in monomer-resolved models}
For both the obtained star polymers and microgels, all pair interactions are modeled with the bead-spring model~\cite{grest1986molecular}.
More precisely, the steric repulsion is modeled via a Weeks-Chandler-Andersen (WCA) potential, acting between two beads $i,j$ with diameters $\sigma_{ij}=(\sigma_i+\sigma_j)/2$, thus also including the core of the star polymers: 
\begin{equation}
\small{V_{\mathrm{WCA}}(r) =
\begin{cases}
4\epsilon \left[ \left( \dfrac{\sigma_{ij}}{r} \right)^{12}
- \left( \dfrac{\sigma_{ij}}{r} \right)^{6} \right] + \epsilon,
& \text{if } r \le 2^{1/6}\sigma_{ij}, \\[8pt]
0, & \text{otherwise.}
\end{cases}}
\label{eq:wca}
\end{equation}
with $\epsilon$ the unit of energy and $r$ the distance between two particles.  In all simulations, the monomer diameter $\sigma$ is the unit of length.
Bonded particles additionally interact with the finitely extensible nonlinear elastic (FENE) potential
\begin{equation}
\small{V_{\mathrm{FENE}}(r)
= -\epsilon k_{\mathrm{F}} R_0^{2}
\ln \!\left[ 1 - \left( \frac{r}{R_0 \sigma_{ij}} \right)^{2} \right]
\quad \text{if } r < R_0 \sigma_{ij}}
\label{eq:FENE}
\end{equation}
with bond stiffness set to $k_{\mathrm{F}} = 15$ and maximum extension $R_0 = 1.5$. 
The temperature-dependent quality of the solvent is mimicked by adding to the bead-spring model a solvophobic potential~\cite{soddemann2001generic}, routinely adopted for PNIPAM–BIS microgels~\cite{ninarello2019modeling}:
\begin{equation}
\footnotesize{V_{\alpha}(r) =
\begin{cases}
-\epsilon \alpha, 
& \text{if } r \le 2^{1/6}\sigma_{ij}, \\[6pt]
\dfrac{1}{2} \alpha \epsilon 
\left[ \cos\!\left( \gamma (r/\sigma_{ij})^{2} + \beta \right) - 1 \right],
& \text{if } 2^{1/6}\sigma_{ij} < r < R_{0}\sigma_{ij}, \\
0, 
& \text{if } r > R_{0}\sigma_{ij}.
\end{cases}}
\label{eq:solvophobic}
\end{equation}
with $\gamma = \pi (2.25 - 2^{1/3})^{-1}$ and $\beta = 2\pi - 2.25\gamma$. 
This attractive potential implicitly accounts for solvent effects and it 
is controlled by the solvophobic parameter $\alpha$, which acts as an effective temperature. 
For $\alpha = 0$, the microgel is fully swollen and no attraction is present, while,  as $\alpha$ increases, the microgel progressively shrinks up to a collapsed state.

\subsection{Simulations of single objects and calculated observables}

Molecular Dynamics simulations of single star polymers and single microgels are performed with the LAMMPS simulation package~\cite{thompson2022lammps} in the NVT ensemble using the stochastic velocity rescaling thermostat~\cite{bussi2007canonical} at a reduced constant temperature $T^* = k_BT/\epsilon = 1$, where $k_B$ is the Boltzmann constant and $T$ the temperature. We perform an initial  equilibration run of $2\times\, 10^6 \delta t$, with $\delta t = 0.002\tau$ where $\tau = \sqrt{m\sigma^2/\epsilon}$ is the unit of time.  Then, a production run is carried out for additional $10\, \times\, 10^6 \delta t$ steps, saving configurations every $5\times 10^4$ time steps. These are used to compute physical observables, such as the radius of gyration, defined as,
\begin{equation}
R_{g} = \left\langle
\left( \frac{1}{N_m} \sum_{i=1}^{N_m} (\vec{r}_{i} - \vec{r}_{\mathrm{cm}})^{2} \right)^{1/2}
\right\rangle ,
\label{eq:Rg}
\end{equation}

the density profile, defined as,
\begin{equation}
\rho(r) = \left\langle
\sum_{i=1}^{N_m} \delta\!\left( \left\lvert \vec{r} - \vec{r}_{i} \right\rvert \right)
\right\rangle ,
\label{eq:density}
\end{equation}
where $r$ is the distance from the center of mass, and the form factor $P(\vec{q}\,)$, defined as:
\begin{equation}
P(\vec{q}\,) = \left\langle
\frac{1}{N_m} \sum_{i=1}^{N_m} \sum_{j=1}^{N_m}
\exp\!\bigl[i\,\vec{q}\cdot(\vec{r}_i - \vec{r}_j)\bigr]
\right\rangle.
\end{equation}

We also compute the hydrodynamic radius, following a protocol recently validated in Ref.~\cite{del2021two} for PNIPAM-BIS microgels in comparison to available experiments. Here, we adapt the method to both star polymers and star-like microgels. In brief, $R_H$ is calculated following Hubbard and Douglas~\cite{hubbard1993hydrodynamic} as,
\begin{equation}
R_{H} =  2 \left[ \int_{0}^{\infty}
\frac{1}{\sqrt{(a^{2}+\theta)(b^{2}+\theta)(c^{2}+\theta)}} \, d\theta
\right]^{-1},
\label{eq:RH}
\end{equation}
with $a$, $b$ and $c$ the principal semiaxes of the effective ellipsoid representing the microgel at each equilibrium configuration. To evaluate this, we adopt two strategies: (i) we construct the convex hull for each equilibrium configuration and take the rigid ellipsoid that has the same gyration tensor of the convex hull~\cite{rovigatti2019connecting}; (ii) we  evaluate the surface mesh using the
$\alpha$-shape method implemented in OVITO~\cite{stukowski2009visualization} with probing radius = $12\,\sigma$.
Representative snapshots for the different macromolecular objects studied in this work are reported in the SI (Fig.~S1) to compare volume estimates obtained from convex hull and surface mesh methods. While the former is preferable for more compact objects since it does not depend on the choice of the probe radius, the latter is more suited to deal with macromolecules whose external surface is highly heterogeneous, such as in the presence of long dangling chains. The probed radius is thus chosen as the minimal one yielding no holes in the surface mesh of the particle. In the limit of very high probe radius, the two methods converge to each other.  In the manuscript, we compare both estimates of $R_H$ showing that they are mostly similar qualitatively, in order to provide robustness to our results. It is important to note that, for large macromolecules such as microgels or star polymers with large number of arms, it would be computationally very demanding to appropriately include hydrodynamic interactions, to reliably evaluate $R_H$. This was done in Ref.~\cite{ruiz2019multi} for stars with $f$ up to 20. We thus use these
previous results to benchmark the present calculations for low-$f$ stars and then extending results for a larger values of $f$.

From the instantaneous variation of the volume $V$ of the effective ellipsoids, we are also able to compute the bulk modulus $K = 1/\kappa_T$, inverse of the isothermal compressibility, as,
\begin{equation}
    K = k_B  T \frac{\langle V\rangle}{\langle V^2 \rangle - \langle V \rangle^2}\, ,
    \label{eq:bulk}
\end{equation}
enabling a direct comparison of the softness of the different examined particles.

\subsection{Calculation of effective potentials}
To calculate the effective potentials, initial configurations of two star polymers or two microgels particles are prepared, randomly oriented with respect to each other, and placed at the center of the simulation box with side $600\sigma$ with periodic boundary conditions, at a relative distance $\mathbf{r_0}$ = $(x_0,0,0)$ from each other. We then wait for any transient to decay and perform a production run for at least $20\, \times\, 10^6 \delta t$ using a Langevin thermostat with friction coefficient $\xi = 1$.

The effective potentials are calculated via the umbrella sampling technique~\cite{blaak2015accurate}, with a harmonic bias of energy $U_{i} = 1/2\, \epsilon\, k_i\, \left[(x-x_0)^2 + y^2 + z^2\right]$ imposed over the natural variables of choice, namely the centers of mass for both microgels and stars or the central cores for stars. Hence, we sample the distances between the variables of choice every $250 \, \delta t$ and run several simulations - windows - spanning $x_0$ over the range of interest, at intervals of $\delta x = 1\, \sigma$. A stiffness of $k_i = 10$ for the bias was utilized at short and intermediate distances, while $k_i = 5$  was used for the  more distant cases. 
For each window, we calculate the biased probability distribution of finding the two particles at the imposed distance, then the harmonic contribution is removed, to calculate the unbiased probability $P(r)$.  Hence, the effective potential is finally calculated as
\begin{equation}
    V_{eff}(r) = - \,k_B \, T\,\ln P(r) + C\, ,
    \label{eq:veff}
\end{equation}
where $C$ is set such that $V(\infty) = 0$.

\section{Experimental methods}
The star-like microgels with  1\% molar percentage of the crosslinker EGDMA are the same used in our previous work~\cite{ballin2025star}. They are synthesized using the same precipitation polymerization method as for standard microgels, just replacing BIS with EGDMA. All details on the synthesis are reported in Ref.~\cite{ballin2025star}.

In this work, we additionally report results of light-scattering measurements made by using a goniometer-based variable multi-angle light scattering (V-MALS) instrument (LS Instruments, Switzerland), equipped with a He–Ne laser (120 mW power, 638 nm wavelength).
The freeze-dried samples were dissolved in Milli-Q water at a concentration of 0.1 mg/ml. Measurements were performed as a function of temperature and scattering angle with the following protocol: a target temperature was set and, once stabilized, the system was allowed to thermally equilibrate for an additional 10 minutes. Subsequently, measurements were carried out across scattering angles ranging from $20^\circ$ to $90^\circ$ in increments of $2^\circ$ (corresponding to a scattering wave vector range of $4.6 \cdot 10^{-3} \, \text{nm}^{-1} \leq q \leq 1.9 \cdot 10^{-2} \, \text{nm}^{-1}$). At each angle, four independent measurements of both the scattered intensity and the intensity autocorrelation function were collected.
The hydrodynamic radius, $R_H$, was determined from the intensity autocorrelation function using a second-order cumulant analysis. This yields the diffusion coefficient, $D$, which is converted to $R_H$ via the Stokes-Einstein relation, $D = k_B T / (6 \pi \eta R_H)$. The reported $R_H$ for each temperature is the average value of measurements at scattering angles between $30^\circ$ and $60^\circ$.

The radius of gyration, $R_g$, was estimated by fitting the angular-dependent scattered intensity, $I(q)$, after normalization by the scattered intensity of toluene, with the Guinier equation:
\begin{equation}
\label{eq:guinier}
    I(q)=I(0)\exp\bigg[-\frac{(qR_g)^2}{3}\bigg]\, ,
\end{equation}
where $I(0)$ is a constant that depends both on the scattering factor of a single particle and on the number of particles in the scattering volume.
To estimate $R_g$, we choose the range of $q$ such that the relation $0.6 \leq qR_g \leq 1.5$ is satisfied. Representative examples of the performed fits in reported in the SI (Fig.~S2).

\begin{figure*}[htb!]
  \centering
\includegraphics[width=0.8\textwidth]{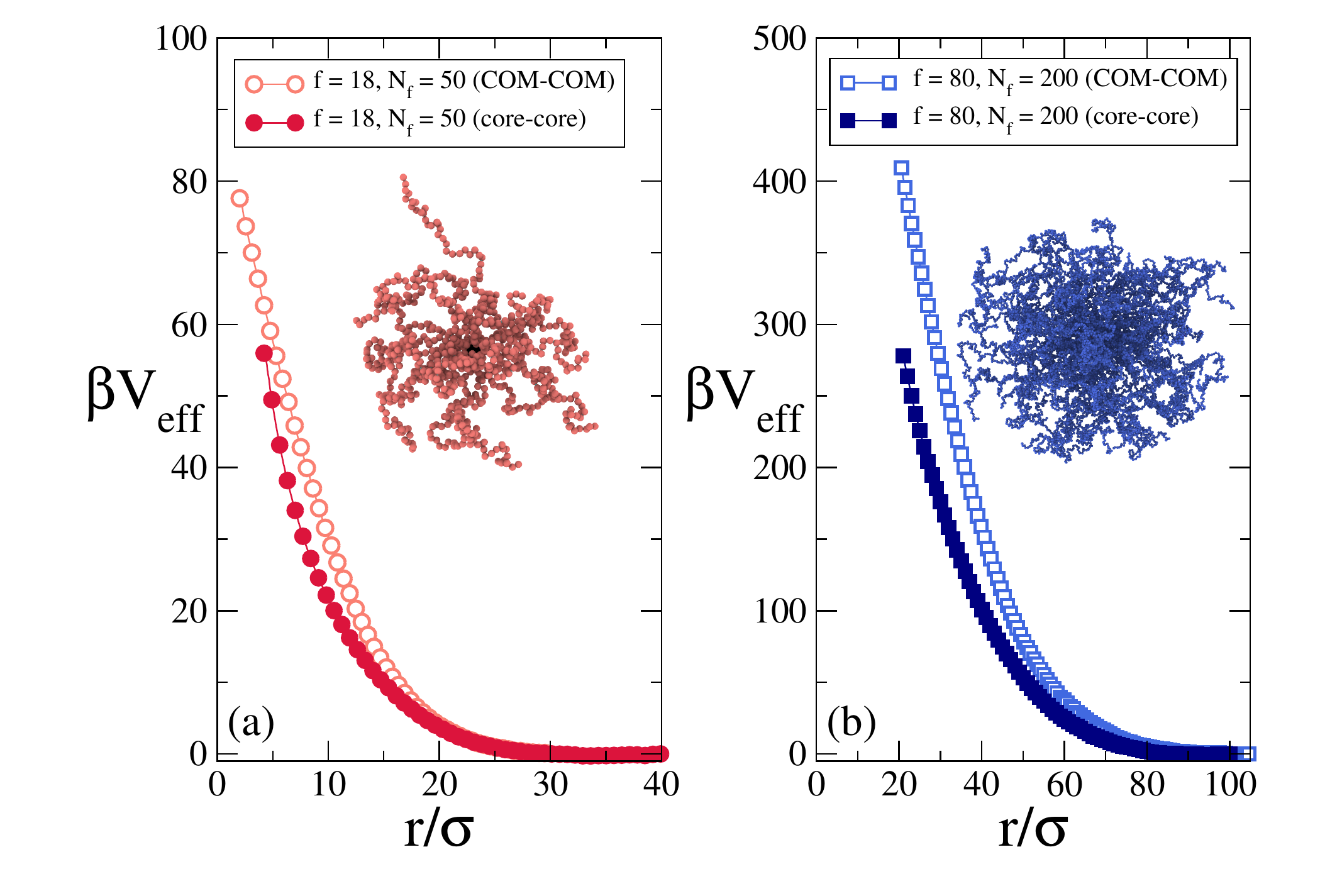}
  \caption{Effective potential $\beta V_{eff}$ between star polymers with (a) $f=18, N_f=50$ and (b) $f=80, N_f=200$ versus both core-core distance $r_{core}$ (full symbols) and COM-COM distances $r_{COM}$ (open symbols). Lines are guides to the eye. The snapshots illustrate a typical equilibrium configuration of the two systems. For the large star with $N_f=200$, the core is hardly visible.
  }
  \label{fig:pot_stars}
\end{figure*}

\section{Results}

\subsection{Effective Potential: Star Polymers}
We start by reporting results for star polymers to make contact with literature and extend previous findings. Fig.~\ref{fig:pot_stars} shows the calculated effective potential for two different kinds of star polymers, namely a low-functionality star with $f=18$ and $N_f=50$ already reported in Ref.~\cite{jusufi1999effective} and a relatively high-functionality star with $f=80$ and $N_f=200$, that was established to correspond to the best representation of c=1\% PNIPAM-EGDMA microgels investigated in Ref.~\cite{ballin2025star}. In doing this, we 
also explore a larger range of $f$ and $N_f$ with respect to available literature results.

Snapshots of both stars are reported in  Fig.~\ref{fig:pot_stars} to visualize the simulated particles. Effective potentials are reported as a function of two possible distances: the most natural variable that expresses a distance between two stars, that is, the distance between the center points of the physical cores $r_{core}$, and a more generic variable that applies to any macromolecules, such as microgels, namely the distance between the centers of mass $r_{COM}$.
As expected, the two potentials are almost identical at large distances, but significant deviations appear at short ones. Indeed, theory predicts a logarithmic divergence for the core-core interactions, which cannot physically overlap on approaching each other, while centers of mass are fictitious points which can also be empty of monomers.
Although we cannot reach even shorter distances within the current simulations,  we see that the slope of the potential between core distances gets steeper as $r_{core}$ decreases, in agreement with expectations, since the two curves should eventually cross. In addition, deviations between core and COM representations 
are more evident when $f$ is larger, due to the increased steepness of the repulsion, leading to stronger deviations between the two representations. To better grasp the microscopic differences between the two potentials, we report the behavior of $r_{COM}$ versus $r_{core}$ in our umbrella sampling simulations and their interplay in the SI (see Figs.~S3-S4 and related discussion).

\begin{figure*}[htb!]
  \centering
  \includegraphics[width=0.7\linewidth]{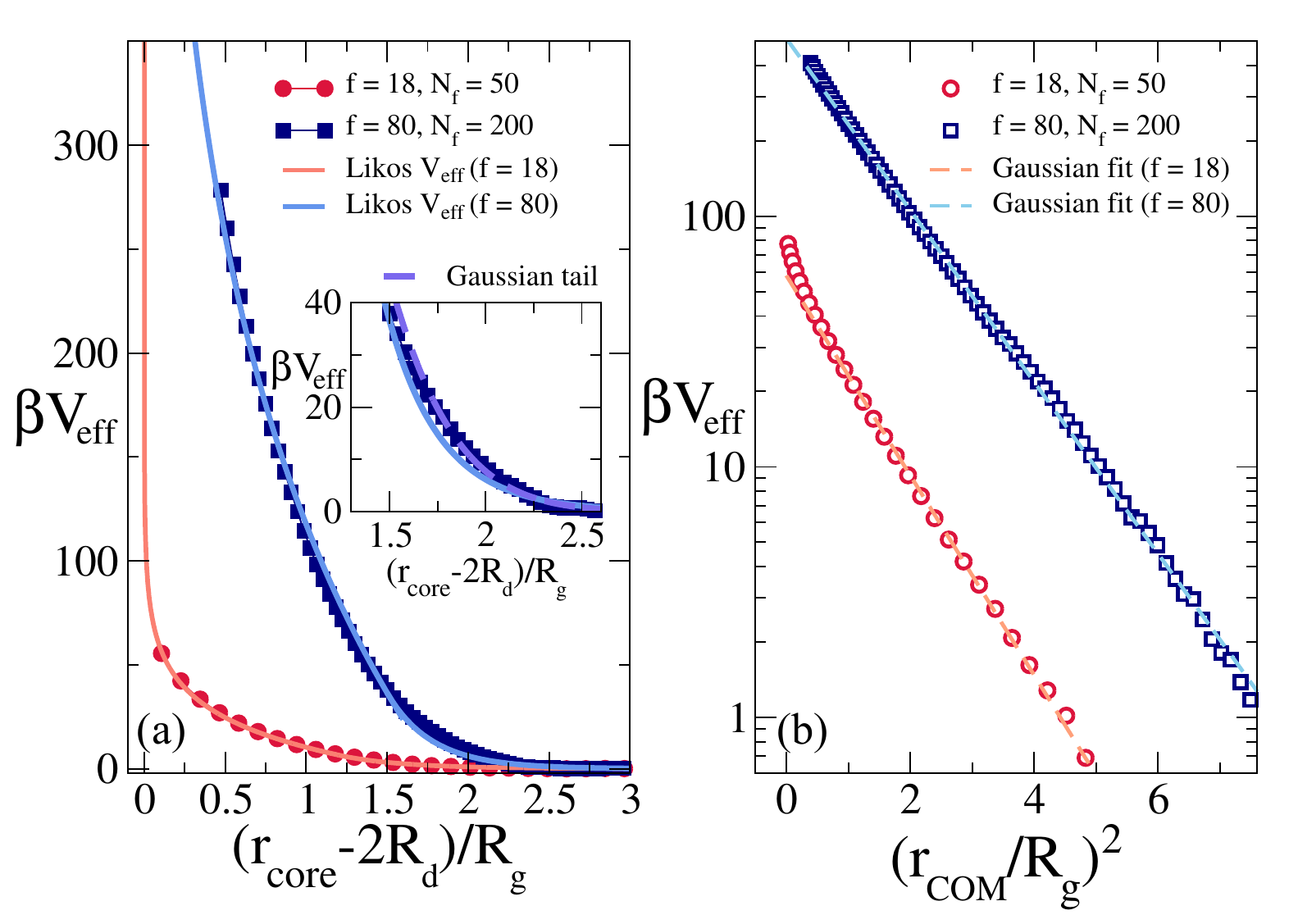}
  \caption{(a): effective potential $V_{eff}$ between star polymers with $f = 50$, $N_f = 50$ and $f = 80$, $N_f = 200$ with $r_{core}$ as the natural variable. The diameter of the first corona of monomers near the core, $2R_d$, is subtracted in order to read the divergence in the origin. The two continuous lines represent the Likos potential fit with a Yukawa tail, as in Eq.~\ref{eq:likos}. The theoretical prefactor $5/18$ is treated as a parameter, reported as $a_0$ in Table~\ref{tab:fit_Likos}, in order to test the adherence to the $f \to \infty$ limit. 
  The inset additionally shows a fit with a Gaussian tail for $f=80$. (b): effective potential $V_{eff}$ between the same star polymers with $r_{COM}$ as the natural variable, normalized by $R_g$.   The dashed lines are Gaussian fits with parameters reported in Table~\ref{tab:fit_gauss}.}
  \label{fig:pot_stars_theory}
\end{figure*}

We compare the calculated effective potentials with theoretical predictions in Fig.~\ref{fig:pot_stars_theory}, for core and COM distances in (a) and (b) panels, respectively.
Starting with the core-core representation, we find excellent agreement between the numerical $V_{eff}$ and the logarithmic ultra-soft potential, that is written in terms of an effective size $\sigma_L$ following Likos and coworkers~\cite{likos1998star}:
\begin{equation}
{V(r)=a_0f^{3/2}\left\{
\begin{array}{ll}
-\ln\!\left(\dfrac{r}{\sigma_L}\right)+\dfrac{1}{1+\sqrt{f}/2}, & r\le \sigma_L,\\[8pt]
\dfrac{\sigma_L/r}{\,1+\sqrt{f}/2}\,\exp\!\left[-\dfrac{\sqrt{f}\,(r-\sigma_L)}{2\sigma_L}\right], & r>\sigma_L ,
\end{array}
\right.}
\label{eq:likos}
\end{equation}
where $a_0=5/18$ from the theory and we use it as a fit parameter to the numerical data.
We find that the logarithmic behavior for $r < \sigma_L $ satisfactorily describes the numerical data with a prefactor $a_0$ in excellent agreement with the predicted value, for both stars as clearly visible in Fig.~\ref{fig:pot_stars_theory}(a) and in Table ~\ref{tab:fit_Likos}. These results validate the Witten and Pincus~\cite{witten1986colloid} and Likos et al~\cite{likos1998star} approaches against numerical results for an unprecedented range of $f$, since in the literature so far only results up to $f=50$ were provided~\cite{jusufi2001effective}.

\begin{table}[h!]
    \centering
    \begin{tabular}{c c c c}
        \hline
        star & \!\!\!\!\! polymers & $a_0$ & $\sigma_L$\\
        \hline
         f=18,& \!\!\!\!\!\! $N_f$=50& 0.2753 & 1.2\\
         f=80,& \!\!\!\!\!\! $N_f$=200&  0.2796& 1.5 \\
         \hline
    \end{tabular}
    \caption{Likos curve fit parameters, as in Eq.~\ref{eq:likos}, for the effective potential between star polymers as a function of $r_{core}$. Note that the divergence has been shifted to $r \sim 0$ using the variable $r=r_{core} - 2R_d$, where $ 2R_d = 2R_c + \sigma$ is the diameter of the first corona of monomers, and the fit parameters $\sigma_L$ are given in units of $(r_{core} - 2R_d)/R_g$. For clarity, we also report $\sigma_L$  in units of $\sigma$: $\sigma_L^{f=18} \sim 17.2 \sigma$, $\sigma_L^{f=80} \sim 54.8\sigma$.}
    \label{tab:fit_Likos}
\end{table}

Regarding the large distance range, we recall that analytical predictions are not available. For this reason, Likos et al~\cite{likos1998star} postulated a crossover to a Yukawa tail, based on the favourable comparison with experimental form factors for $f=50$.
This tail is however incompatible with the Gaussian behavior observed for linear chains ($f=2$), which led to the ansatz that for $f<10$ the potential should cross over to a Gaussian form~\cite{jusufi2001effective}. 
This problem was later revisited by Hsu and Grassberger~\cite{hsu2004effective}, who claimed a better description of the Gaussian form at all $f$.

Here, we show in Fig.~\ref{fig:pot_stars_theory}(a) that the Yukawa representation is  qualitatively satisfactory for $f=18$ but less accurate for $f=80$ (see inset for a magnification at large distances). Indeed, it seems that a Gaussian tail, also shown in the inset of Fig.~\ref{fig:pot_stars_theory}(a), reproduces the numerical data better. 

Focusing instead on the COM representation, we report $V_{eff}(r_{COM})$ in Fig.~\ref{fig:pot_stars_theory}(b) for both stars, finding that a Gaussian description~\cite{hsu2004effective},
 \begin{equation}
V_{eff}\,(r) = c_f e^{-d_f (r/R_g)^2}\, ,
\label{eq:gaussian}
\end{equation}

successfully applies to almost all probed distances, except when the cores are very close to each other, with $c_f$ setting the energy scale and $d_f$ being related to the inverse of the standard deviation.
The corresponding fit parameters are reported in Table~\ref{tab:fit_gauss}, yet a direct comparison to the lattice results of Hsu and Grassberger is not easily possible. Notwithstanding this, the robustness of the Gaussian dependence on $r_{COM}$ appears to be quite remarkable. It is however reassuring that at large distances the differences between the two representations of the potential are minor, thus not substantially affecting results, as also shown by  the numerous validations of the Likos potential against experiments~\cite{likos1998star,stiakakis2002polymer,zaccarelli2005tailoring,ruiz2018crystal}. 

\begin{table}[h!]
    \centering
    \begin{tabular}{c c c c}
        \hline
        star & \!\!\!\!\! polymers & $d_f$ & $c_f$\\
       \hline
         f=18, &\!\!\!\!\! $N_f$=50& 0.92 & 56.6\\
         f=80, & \!\!\!\!\! $N_f$=200&  0.78& 491\\
  \hline
    \end{tabular}
    \caption{Gaussian fit parameters for the effective potential between star polymers as a function of $r_{COM}$.}
    \label{tab:fit_gauss}
\end{table}

\subsection{Effective Potential: Standard and Star-like Microgels}

\begin{figure*}[t!]
  \centering
  \includegraphics[width=0.7\linewidth]{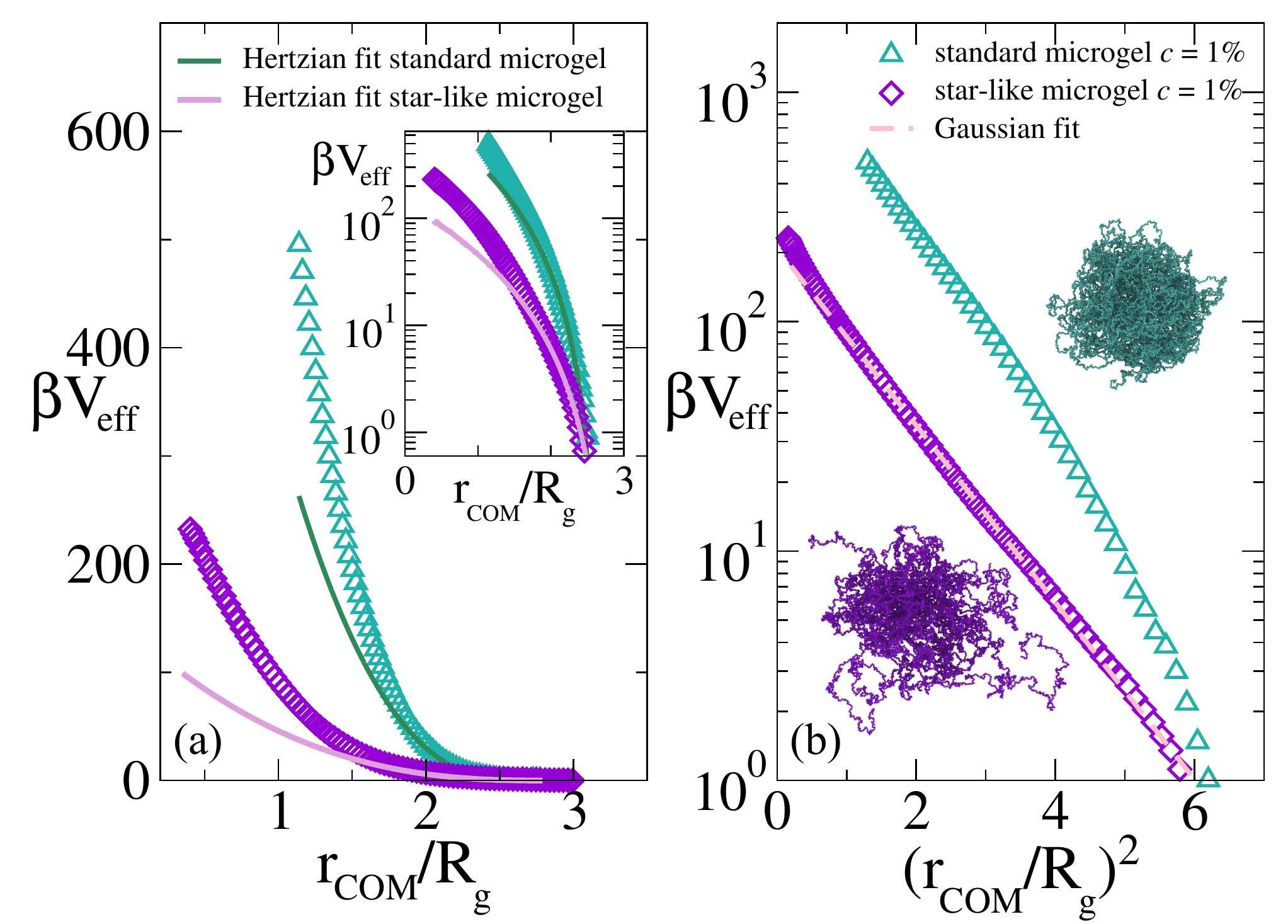}
  \caption{ Left panel: effective potential $V_{eff}$ between star-like microgels and standard microgels at a crosslinker concentration $c = 1\%$. Hertzian fits are shown for both systems; the tail is highlighted in the log-scale inset. Right panel: the same potentials are shown on a Gaussian scale, where Gaussian profiles appear as straight lines. The Gaussian fit for the star-like microgel is shown as a dashed line. Fit parameters are reported in Tab.\ref{tab:fit_mgels}.}
  \label{fig:pot_mgels}
\end{figure*}

We now proceed by reporting the calculated effective potential between microgels, comparing star-like and standard ones at the same crosslinker concentration $c=1\%$. Results are reported in Fig.~\ref{fig:pot_mgels} as a function of the natural variable $r_{COM}$. We first observe in Fig.~\ref{fig:pot_mgels}(a) that $V_{eff}$ is much more repulsive for standard microgels as compared to star-like ones, when they are compared at the same relative distance. This is due to the fact that the corona of the star-like microgel is much more extended and less compact, as visible from the snapshots also reported in the Figure, which justifies the weaker interactions.

It is well-established that standard microgels under dilute conditions are a good model system for elastic spheres, interacting with the Hertzian  potential, defined as
\begin{equation}
V(r)\, =\, 
\begin{cases}
        \epsilon_H\, {\left(1 - \cfrac{r}{\sigma_H} \right)}^{\frac{5}{2}}\, , & \text{if }\, r < \sigma_H\, , \\
        0\, , & \text{otherwise}\, ,
\end{cases}
\end{equation}
with an energy scale $\epsilon_H$, which depend on the elastic properties of the microgel, and an interaction range $\sigma_H$, usually found to be coinciding with the hydrodynamic diameter~\cite{paloli2013fluid}. Here we confirm these results, as shown in Fig.~\ref{fig:pot_mgels}(a), which also reports a Hertzian fit to the numerical results, whose parameters are provided in Table~\ref{tab:fit_mgels}. In agreement with previous results~\cite{rovigatti2019connecting}, the Hertzian description holds up to a moderate range of distances, below which the calculated potential is more repulsive, due to the heterogeneous core-corona structure of the microgel, which increases the internal elastic response of the particles at strong compression~\cite{bergman2018new}. A similar description also applies to the star-like microgel, albeit the deviations from the Hertzian form appear to be more pronounced. 
We argue that the Hertzian behavior for the star-like microgels effective potential is only apparent and that instead a Gaussian description is valid for most probed distances. This is shown in Fig.
~\ref{fig:pot_mgels}(b), where a Gaussian fit is seen to apply to the numerical data for several orders of magnitude. The same does not clearly hold for the regular microgel, which shows clear deviations from a Gaussian shape. Interestingly, the resulting Gaussian fit parameters, especially  $d_f$, also reported in Table~\ref{tab:fit_mgels}, are rather similar to those found for star polymers, reported in Table~\ref{tab:fit_gauss}.

\begin{table}[h!]
    \centering
    \begin{tabular}{c c c c c c}
        \hline
        microgel &\!\!\!\!\!\!\!type & $\epsilon_H$ & $\sigma_H$ & $c_f$ & $d_f$\\
        \hline
         \emph{star-like} & \!\!\! $c = 1\%$& 138.5 & 103.3 & 209 & 0.89\\
         \emph{standard} & \!\!\! $c = 1\%$&  1092& 73.6 & - & -\\ 
         \hline
    \end{tabular}
    \caption{ 
    Fit parameters for star-like and standard microgels effective potential. Both Hertzian and Gaussian parameters are reported. Note that $\sigma_H$ is given in units of the monomer size $\sigma$. As a reference, $R_g$ is found to be $37.0\sigma$ and $28.1\sigma$  for star-like and standard microgels, respectively.}
    \label{tab:fit_mgels}
\end{table}

\subsection{Comparison between Star Polymers and Star-like Microgels}
\begin{figure*}[htb!]
  \centering
  \includegraphics[width=1\linewidth]{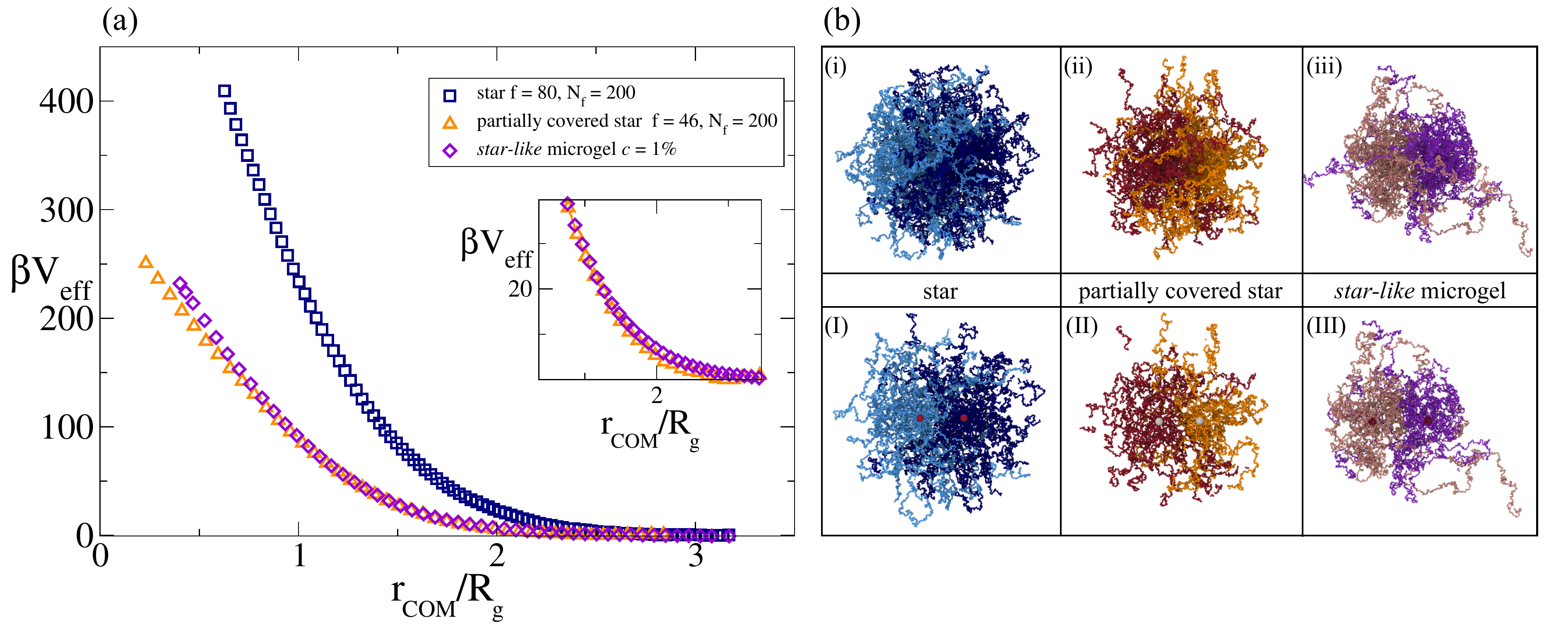}
  \caption{(a) Effective potential $\beta V_{eff}$ between $c=1\%$ star-like microgel,  between star polymers with $f=80$, $N_f=200$, $R_c = 2.2\, \sigma$ and between partially covered stars with $f=46$,$N_f=200$, $R_c = 2.2 \, \sigma$, as a function of $r_{COM}$ rescaled by the respective gyration radii. The inset enlarges the short distance behavior; (b) Snapshots between two of each macromolecus at a distance $r_{COM}/R_g \sim 0.6$  to visually explain the differences observed in the effective potentials. The top row shows the full macromolecules, while the bottom one reports the corresponding slices. Note that the cores of the stars as well as the crosslinkers  of the star-like microgels have a different color. }
  \label{fig:pot_comparison_star_mgel}
\end{figure*}

After having discussed star polymers and microgels separately, we report a direct comparison between the effective potential for star-like microgels and for star polymers with $f = 80$, $N_f = 200$ as a function of $r_{COM}$. Given their similar structure, already documented in Ref.~\cite{ballin2025star},  we would also expect their interaction potentials to be similar.  However, although sharing a similar Gaussian profile given by the comparable values of $d_f$ (see Table~\ref{tab:fit_gauss}), we find that $V_{eff}$ between star polymers is much more repulsive than the corresponding one for star-like microgels, as shown in Fig.~\ref{fig:pot_comparison_star_mgel}(a). This originates from the much stronger local crowding of chain monomers near the  cores of the stars, as compared to the same taking place around the microgel crosslinkers. This effect is clearly visible in the snapshots reported in  Fig.~\ref{fig:pot_comparison_star_mgel}(b), taken at a comparable distance.
Clearly, the microgels are far less dense in monomers in their inner part, making them to interact with a softer repulsion. Given this observation, which stems from the full coverage of the cores of the stars, as opposed to the looser structure around the crosslinkers of the microgel, we reduce the number of arms of the star polymer, while  keeping $R_c$ fixed. We thus build a star polymer with a partially covered surface of the core, choosing $f=46$ which corresponds to an effective coverage fraction $\gamma \sim 0.57$. For this specific value, we remarkably find an effective potential that is virtually indistinguishable from that between star-like microgels, as also shown in Fig.~\ref{fig:pot_comparison_star_mgel}(a) and in the relative inset. Indeed, the snapshots between the partially covered stars are very similar to those between star-like microgels. Hence, the local monomer density around the cores or the crosslinkers appears to be the main parameter distinguishing pure stars from star-like microgels.

\subsection{About star polymers with reduced core surface coverage}

\begin{figure}[h!]
    \centering
    \includegraphics[width=1\linewidth]{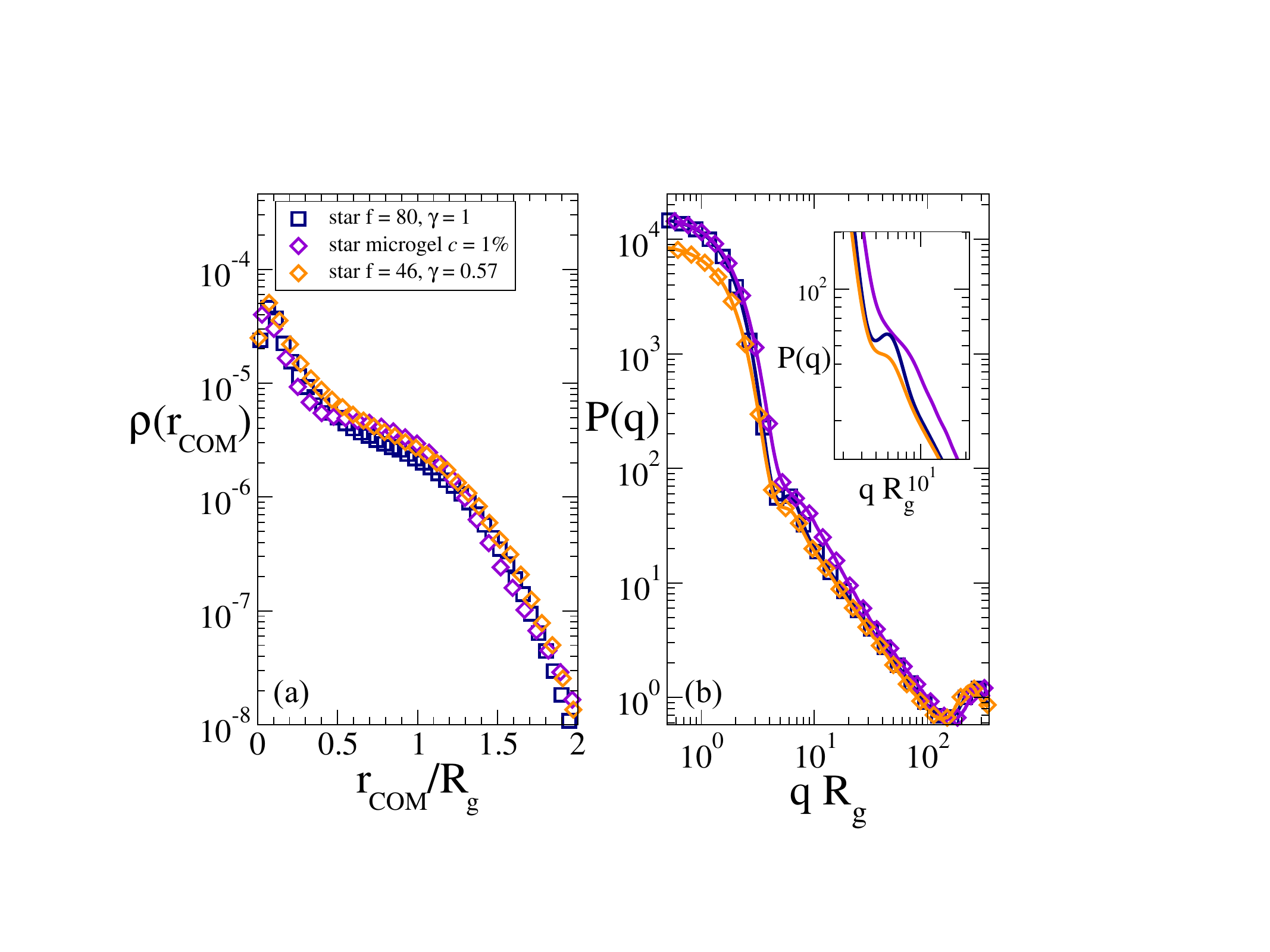}
    \caption{(a) Density profiles $\rho(r_{COM})$ and (b) form factors $P(q)$ for the star polymer with $f = 80$, $N_f = 200$, for the \emph{star-like} microgel with $c = 1 \%$, and for the partially covered star with $f = 46$, $N_f = 200$. Data in panel (a) are normalized to yield a volume integral equal to 1, while in (b) they are only scaled by the respective gyration radii. The inset shows a zoom-in of the peak profiles.}
    \label{fig:dens_profiles}
\end{figure}

The effect of the core surface coverage on the physical properties of a star has not been investigated in detail before. We thus report here the variation of the structure, quantified by the density profile of the particles, calculated with respect to the centers of the mass, in Fig.~\ref{fig:dens_profiles}. We see that all profiles, namely that for a pure star with $f=80$ and full coverage ($R_c = 2.2\sigma$), for a partially covered star with $f=46$ ($R_c = 2.2\sigma$), and for the star-like microgels, are all very similar. As expected, also the form factors are similar, although the star with full coverage clearly has a more pronounced peak, as visible in the inset. This is indicative of a more crowded zone near the core, given its high functionality, that leads to the discussed more repulsive interactions. On the other hand, the partially covered star and the star-like microgel share a similar peak profile. 

\begin{figure}[!htbp]
  \centering
  \includegraphics[width=1\linewidth]{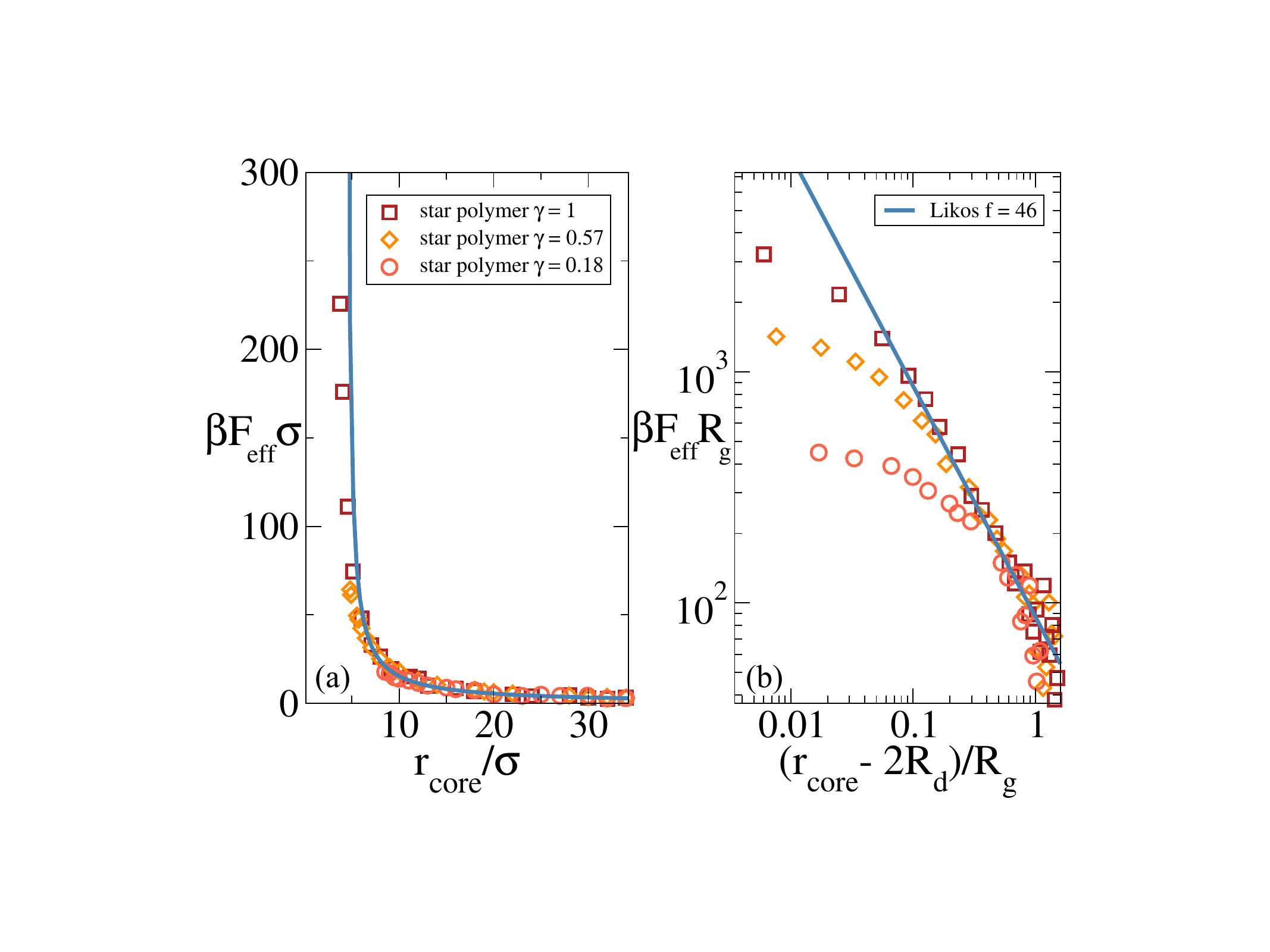}
  \caption{Effective mean force $\beta F_{eff}$ for three star polymers with fixed $f = 46$, $N_f = 200$ and different coverage fractions $\gamma$, achieved by changing $R_c$ as explained in the text. In panel (a) data are shown as a function of  $r_{core}$, while in panel (b) this is shifted by $2R_d$ similar to previous studies~\cite{likos1998star}. The line in both plots represents the negative gradient of the Likos potential in Eq.~\ref{eq:likos}, i.e. the force. In panel (a), this is set to diverge for the fully covered case.}
  \label{fig:star_full_vs_partial_shiftRd}
\end{figure}

So far, we have adjusted the coverage fraction $\gamma$ by keeping the core size fixed while varying the star polymer functionality $f$, thus probing the dependence $\gamma=\gamma(f)$. It is also interesting to investigate whether, at fixed functionality $f$, the core size has a significant effect on the star polymer interactions. To this end, we consider three systems in which $f$ and $N_f$ are fixed, while $\gamma = \gamma(R_c)$ is tuned by changing the core radius $R_c$. In particular, we use $R_c = 1.7\,\sigma$ ($\gamma \sim 1$), $R_c = 2.2\,\sigma$ ($\gamma \sim 0.57$), and $R_c = 4\,\sigma$ ($\gamma \sim 0.18$). For these systems, we calculate the two-body mean force as a function of the core–core distance, as commonly done for star polymers~\cite{huissmann2009star}.
The results are plotted in Fig.~\ref{fig:star_full_vs_partial_shiftRd}, showing that 
over a wide range of distances, the mean-force curves as a function of $r_{core}$  are nearly indistinguishable among each other despite the change of coverage (see panel (a)).

This weak sensitivity to $R_c$ is expected in the regime $N_f \gg f^{1/2}$, where the arms are long enough that the dominant contribution comes from the swollen corona and the interaction is set mainly by the overall number of arms (and the outer length scale), rather than by microscopic details of the grafting pattern. Small but systematic differences only emerge at short separations, where finite-core effects become relevant and the mapping onto the asymptotic star potential deteriorates. To visualize this, we plot in panel (b) the same data as a function of $(r_{\rm core}-2R_d)/R_g$ in a log-log representation.
As expected, deviations from the logarithmic divergence increase with increasing core radius. 
This trend is consistent with a reduced number of chains effectively participating in the contact region when $\gamma < 1$, which lowers local crowding, and thus the entropic penalty, at a given core separation.

These results show that the effect of the coverage fraction is only relevant close to the core, thus explaining why a star-like microgel has similar interactions to those of a pure star, although less repulsive (see Fig.~4). Indeed, the coverage fraction of the star-like microgel is lower than that of the stars, but not so much for low $c$, where the crosslinkers-rich core is small. 
Instead, for PNIPAM-EGDMA microgels at higher $c$, where the core size becomes significant as shown in Ref. ~\cite{ballin2025star}, the star-like behavior is already lost in the overall structure, so it 
will be interesting to see what happens in terms of interactions.

\subsection{Properties across the Volume Phase Transition}

To provide a comprehensive characterization of the different macromolecules and to explore their behavior upon increasing temperature, we also carry out simulations in the presence of the solvophobic potential (Eq.~\ref{eq:solvophobic}) varying the parameter $\alpha$, which plays the role of an effective temperature. The VPT transition for PNIPAM-based microgels occurs for $\alpha\sim 0.65$. However, Ref.~\cite{ballin2025star} has shown
that the VPT occurs much more sharply for star-like microgels than for regular ones, as also captured by our numerical model. This is due to the strong decoupling between the arms and the core that occurs in star-like objects, as opposed to crosslinked microgels, whose greater connectivity smoothens the transition.

We characterize the different systems by looking at two observables as a function of temperature, namely the ratio between gyration and hydrodynamic radius $R_g/R_H$, also in comparison to new experimental measurements, as well as the bulk modulus which is a measure of the softness of the particles.

\subsubsection{Ratio between gyration and hydrodynamic radius }
\label{sec:Rg/Rh}
The ratio between gyration and hydrodynamic radius is often used to characterize the mass distribution within a macromolecule\cite{del2021two}. In particular, while $R_g$ quantifies the polymeric distribution  around the center of mass, the hydrodynamic radius is more sensitive  to the presence of the outer chains, contributing to the diffusion of the particle. Hence, while for homogeneous spheres, $R_g/R_H$ is found to be $\sqrt{3/5}$, for star polymers with low arm number values greater than 1 have been reported~\cite{roovers1972preparation,roovers1983analysis,bauer1989chain,ohno2002scaling}. Regular microgels have been reported to have a value of this ratio close to 0.5-0.6, tending to the HS value as temperature increases and the particle collapses~\cite{del2021two,elancheliyan2022role}. It is then important to estimate this ratio also for star-like microgels, in order to further assess their similarity to star polymers. 

\begin{figure}[!htbp]
    \centering
    \includegraphics[width=1\linewidth]{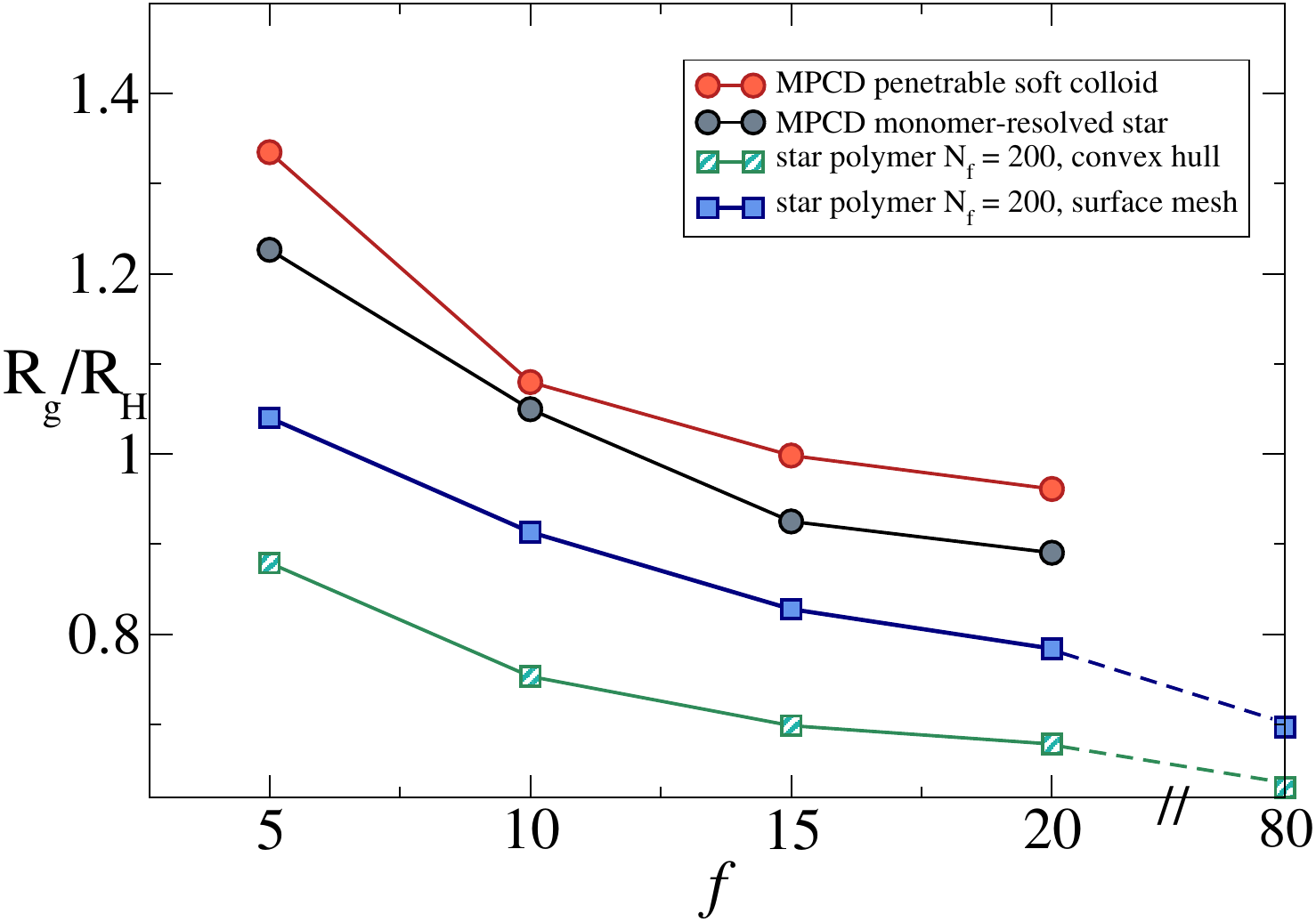}
    \caption{Ratio between gyration and hydrodynamic radius in good solvent ($\alpha = 0$) for star polymers as a function of $f$ with the two methodologies (convex hull vs surface mesh) to calculate $R_H$. Comparison with MPCD results from Ref.~\cite{ruiz2019multi}.}
    \label{fig:single/Rg_RH_alfa0}
\end{figure}

To do so, it is important to first establish a meaningful way to estimate the hydrodynamic radius in simulations without explicit hydrodynamics, to avoid the high computational cost associated with explicitly accounting them in such complex macromolecules. We thus resort to a method recently reported for regular microgels, as described in Methods. To validate this procedure in the case of star polymers and to extend current calculations to large values of $f$, needed to compare to star-like microgels, we first report $R_g / R_H$ in good solvent for star polymers with different functionalities in Fig.~\ref{fig:single/Rg_RH_alfa0}. The numerical calculations are performed using two different approximations to estimate the volume of the particle, the convex hull and the surface mesh as explained in Methods and also illustrated in the SI (see Fig.~S1). Both approaches yield qualitatively similar results, where the ratio decreases with  functionality, also in agreement with available numerical estimates for low $f$ including explicit hydrodynamic interactions (HI)~\cite{ruiz2019multi}. In this study, two approaches based on Multi-Particle Collision Dynamics (MPCD)~\cite{gompper2009multi} were compared: a monomer-resolved study vs a penetrable soft colloid model, both yielding qualitatively similar results.
It is clear that, while for small $f$ the ratio $R_g / R_H$ strongly exceeds 1 in the presence of HI, it decreases upon increasing number of arms. This is expected, since, for $f\to \infty$, one should recover the HS limit.  We find that the present calculations based on the surface mesh are quantitatively closer to the monomer-resolved results in the presence of HI, and although our estimate is only qualitative,
the trend with $f$ is robust and scales similarly to the HI results~\cite{ruiz2019multi}.

\begin{figure}[!htbp]
    \centering
    \includegraphics[width=1\linewidth]{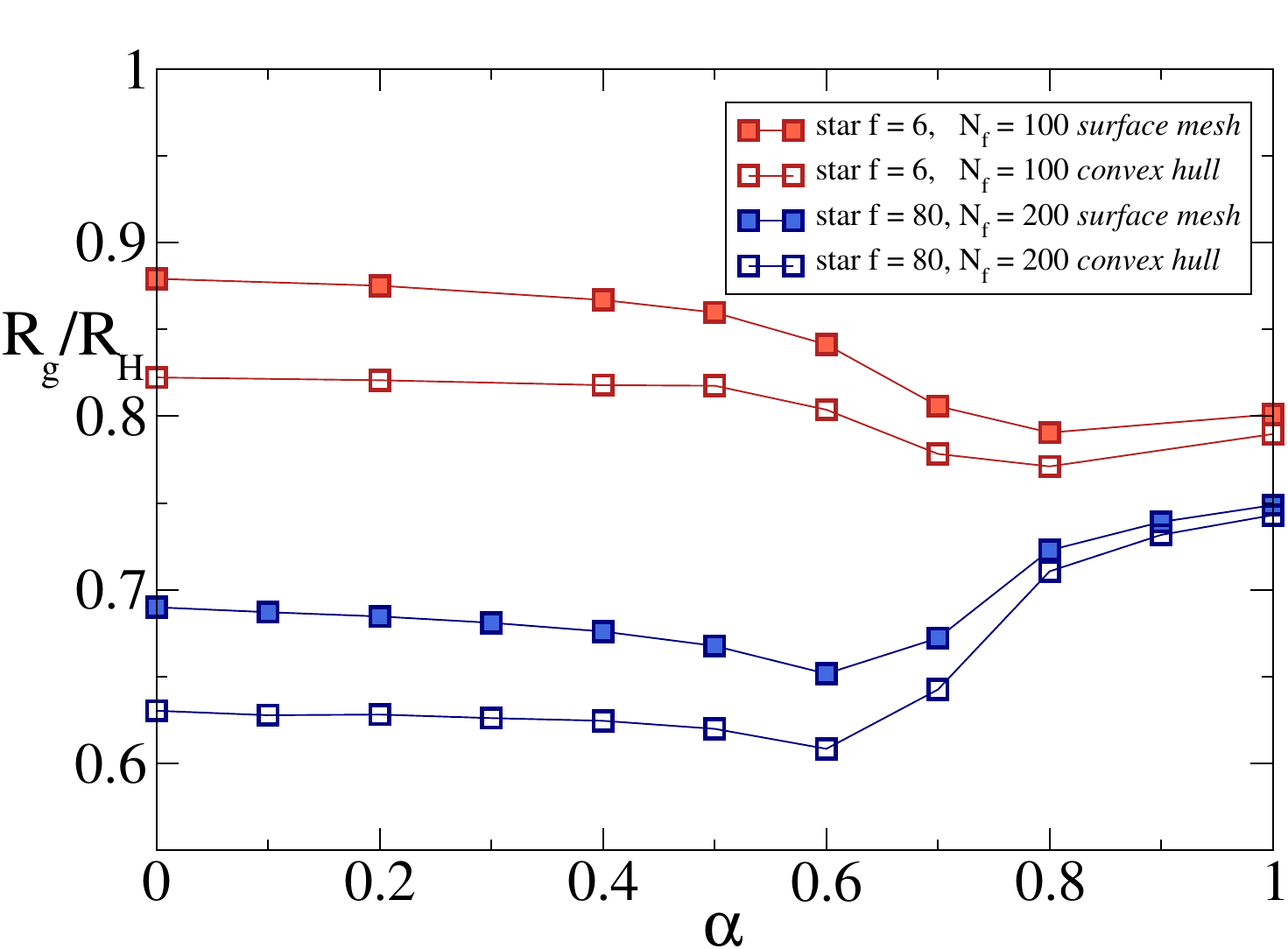}
    \caption{$R_g/R_H$ for two star polymers, one with low functionality $f = 6, N_f = 100$, one with high functionality $f = 80, N_f = 200$. The graph shows how the ratio $R_g/R_H$ decreases going from good solvent to bad solvent for the low functionality star, increases for $f = 80$, exhibiting a minimum at the VPT transition. The surface mesh and convex hull curves yeld the same qualitative behavior.
    }
    \label{fig:single/Rg-RH_stars}
\end{figure}

Having established the reliability of our calculations, we next report the behaviour of $R_g/R_H$ as a function of the solvophobic parameter $\alpha$. First of all, we focus on star polymers and find that a very different trend with $\alpha$ takes place for low versus high  number of arms, as shown in Fig.~\ref{fig:single/Rg-RH_stars}. Indeed, while $R_g/R_H$ decreases monotonically with $\alpha$ for low-$f$ stars, for high-$f$ star polymers the ratio is found to first decrease and then increase again, exhibiting a characteristic minimum associated with the VPT transition, already observed in regular microgels~\cite{del2021two,elancheliyan2022role}. Interestingly, in those cases, the minimum was attributed to the presence of charges on the external microgel corona, which delays the collapse of the outer chains with respect to the core and was not observed in simulations of neutral microgels, albeit with a larger crosslinker concentration and only for the convex hull calculation~\cite{del2021two}. Here, we find that the minimum exists for neutral stars, independently of the method used. This is probably due to the higher degree of heterogeneity of the particles, whose outer chains may undergo deswelling earlier than the inner ones. 

\begin{figure}[!htbp]
    \centering
    \includegraphics[width=1\linewidth]{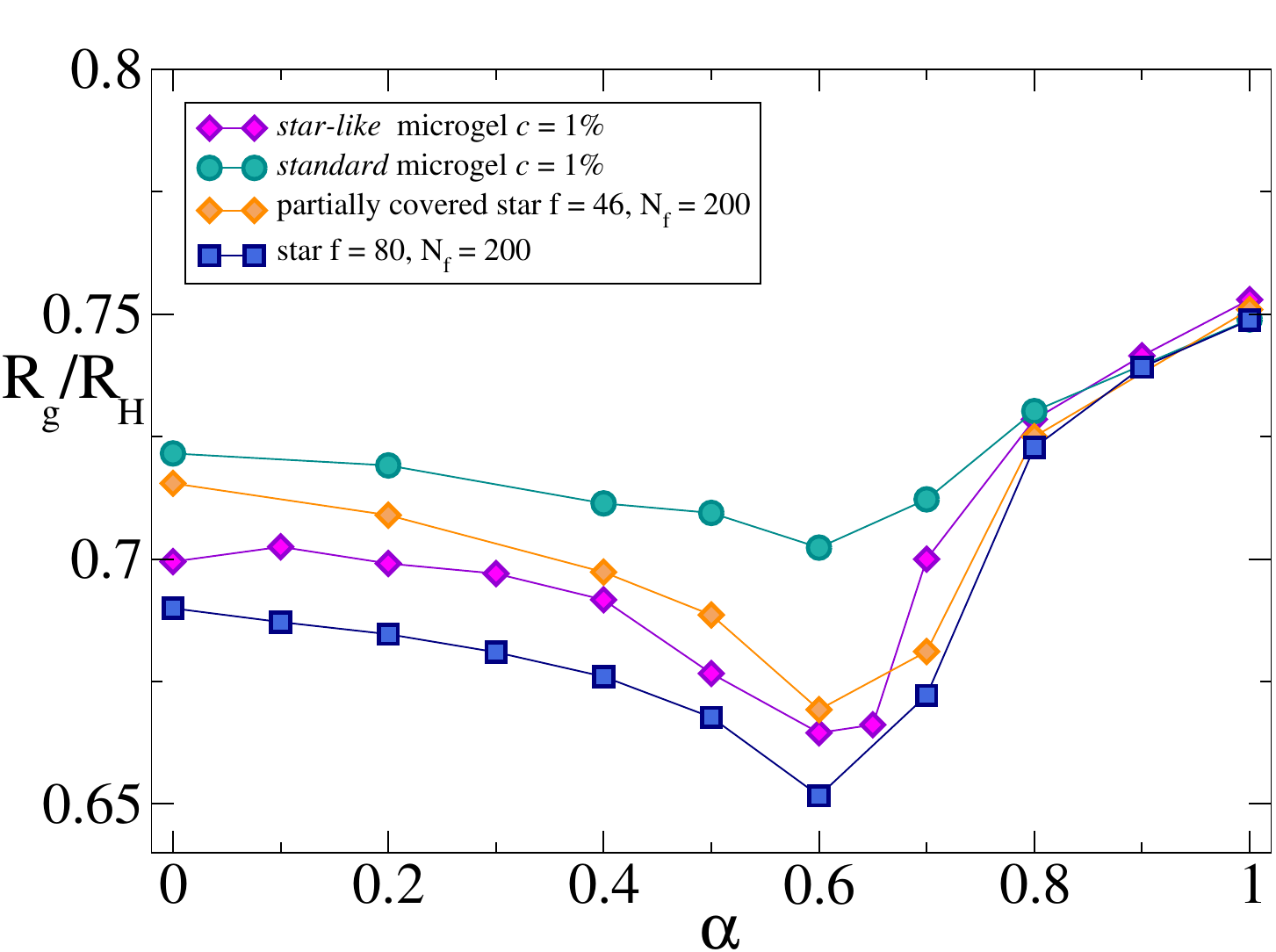}
    \caption{Ratio $R_g/R_H$ for a \emph{star-like} microgel with $c$=1\%, for a \emph{standard} microgel also with $c$=1\%,  for the fully covered star ($f = 80$,$N_f = 200$) and for a partially covered star with $f = 46$ and $N_f = 200$ ($\gamma \sim 0.57$). The latter is the one that best reporduces the effective potential of the \emph{star-like} microgel (see Fig.~\ref{fig:pot_mgels}). The ratio is estimated in all cases using the surface mesh method for the hydrodynamic radius.}
    \label{fig:single/Rg-RH_mgels}
\end{figure}

\begin{figure}[!htbp]
    \centering
    \includegraphics[width=1\linewidth]{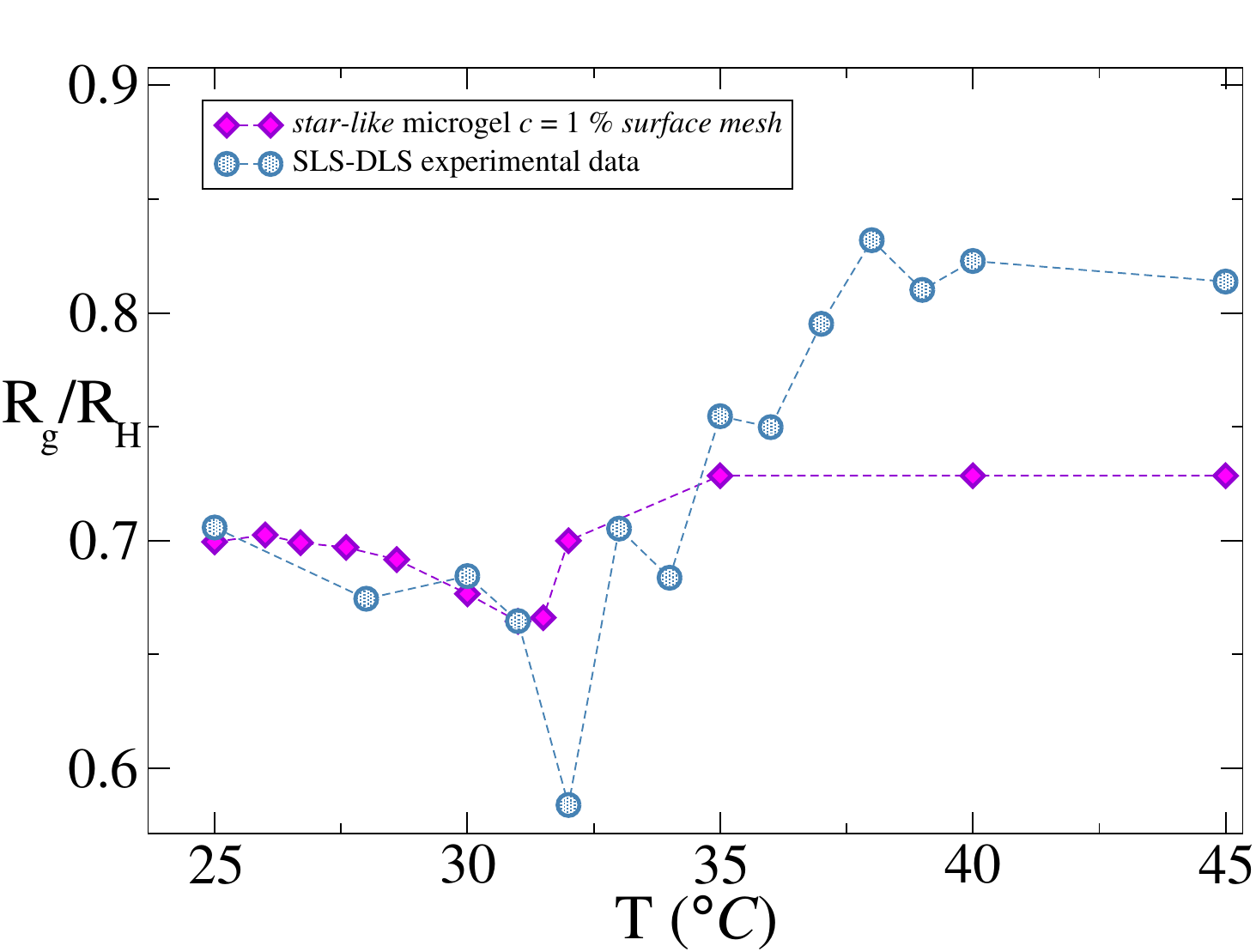}
    \caption{Comparison for $R_g/R_H$ between \emph{in-silico} \emph{star-like} microgel and SLS-DLS experimental data. Dashed lines are to guide the eye.
    }
    \label{fig:single/Rg-RH_experiments}
\end{figure}

Next, we compare the behavior of $R_g/R_H$ for the star polymer with $f = 80$ to the corresponding one for the star-like microgel with $c = 1\%$. In order to have a full picture, we also show the ratio calculated for a regular microgel, again with $c=1\%$, and for the partially covered star with $f = 46$. We find that, while all data are generally similar, star-like microgel is the one showing the sharpest growth as $\alpha$ is further increased. This is especially true when compared to the regular microgels, reflecting the sharpness of the VPT of PNIPAM-EGDMA star-like microgels already observed by DLS~\cite{ballin2025star}. In addition, the fully covered  and the partially covered stars seems to bracket the values of the ratio  $R_g/R_H$ for the star-like microgel. All three star-like systems display a very pronounced minimum in $R_g/R_H$, probably due to the cooperative nature of the arms undergoing the the VPT as compared to the fixed core. In contrast, for the standard microgel the minimum is barely detected.  Note that, with respect to the data reported in Ref.~\cite{del2021two}, where the minimum was not found for neutral microgels, here we find it for a slightly lower $c$ and using the surface mesh method. Indeed, the minimum tends to disappear when using the convex hull for standard microgels, as shown in the SI (Fig.~S7). 

To verify these predictions, we directly compare $R_g/R_H$ for the star-like microgel with experimental results, obtained by SLS and DLS experiments on PNIPAM-EGDMA microgels with $c = 1\%$.
The experimental swelling curves for $R_g$ (SLS) and  $R_H$ (DLS) are also shown in the SI (Fig.~S5), together with the corresponding ones obtained by simulations (Fig.~S6). 
We find that the direct comparison of experiments and simulations is in qualitative agreement.  As expected, the minimum of $R_g/R_H$  is much more pronounced in experiments, due to the presence of electrostatic effects introduced by the ionic initiator~\cite{del2021two}, not considered for simplicity in the simulations. It is confirmed also in experiments that for large enough number of arms the ratio does not exceed 1, at swollen conditions, and tends to the HS value at large temperatures. Again, the sharpness of the VPT is evident from the experimental data.

\subsubsection{Bulk modulus }
\label{sec:bulk}
\begin{figure}[!htbp]
  \centering
  \includegraphics[width=1\linewidth]{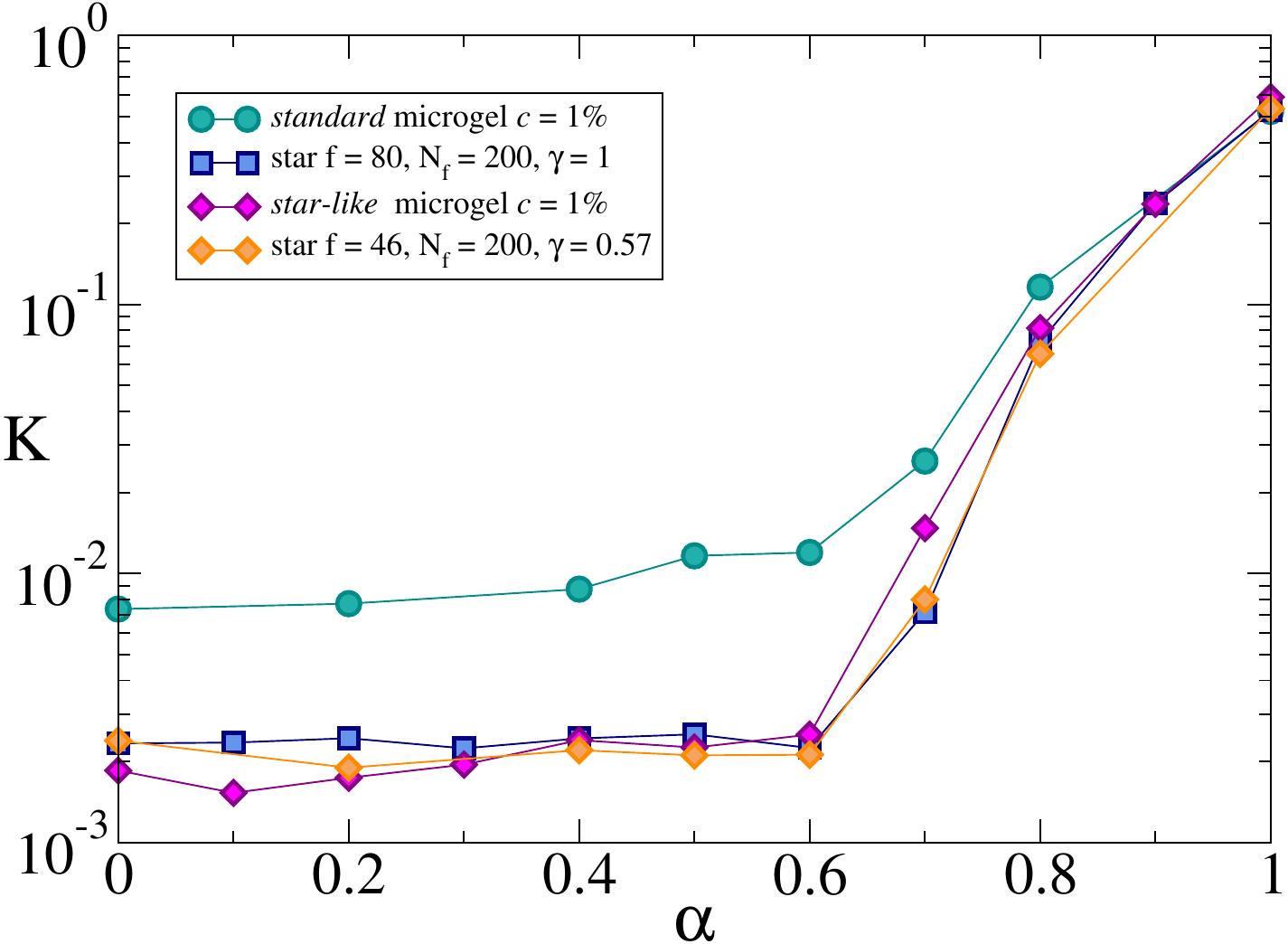}
  \caption{Bulk modulus $K$ in units of $k_BT/\sigma^3$ as a function of effective temperature $\alpha$, obtained using the surface mesh method to estimate the volume.}
  \label{fig:bulk_modulus}
\end{figure}

Finally, to provide an estimate of the softness of the particles we also calculate their bulk modulus $K$ as a function of effective temperature. This is simply estimated from the volume fluctuations, as done in previous works~\cite{rovigatti2019connecting}.This is reported in Fig.~\ref{fig:bulk_modulus},  which shows that the star-like microgel, as well as the partially covered star, displays a bulk modulus almost one order of magnitude lower than regular microgels with the same degree of crosslinking. No minimum in $K$ is detected  at the VPT, for any system. Finally at large $\alpha$, under collapsed conditions, all moduli tend to coincide. Similar results are obtained when using the the convex hull method (see Fig.~S8), showing qualitatively similar results. 
Given the large softness of star-like microgels, it is instructive to compare the calculated value of $K$ for $\alpha=0$ with the corresponding one estimated for ultra-low-crosslinked (ULC) microgels of comparable numerical size~\cite{marin2025unexpected}. ULC microgels are synthesized in the absence of crosslinkers and are considered to be the softest existing type of microgels. Their bulk modulus $\sim 10^{-3}$ is thus confirmed to be slightly lower than for star-like microgels, but only by a factor $\sim$ 2, as compared to the order of magnitude difference found with respect to regular microgels.

\section{Conclusions}
In this work we reported extensive numerical simulations of star-like microgels with $c=1\%$, easily realizable in the laboratory simply by substituting BIS with EGDMA crosslinkers in the chemical synthesis~\cite{ballin2025star},
and compared their structural properties and their effective potentials to those of star polymers and standard microgels. Through a detailed analysis of all these quantities, we have demonstrated that star-like microgels with low degree of crosslinkings are a different class of soft colloids with respect to standard microgels. In particular, they are much softer, as demonstrated by their elastic moduli roughly one order of magnitude lower, and they display a Gaussian effective potential in a wide range of relative distances, as opposed to the Hertzian-like one of regular microgels. The latter feature appears to be also very different from recent results for ULC microgels, which seem not to obey any effective pair potential due to the their extreme deformability~\cite{marin2025unexpected}. Thus, despite a similar value of the bulk modulus, star-like microgels appear to be also distinguished from ULC microgels.

On the other hand, the characteristics evidenced in this work actually establish a deep connection between star-like microgels and star polymers, since they share the same functional form of the effective potential, as well as a similar behavior of the elastic moduli and of the ratio between gyration and hydrodynamic radii. However, we also found that ideal stars with full core coverage experience a much more repulsive interaction than star-like microgels, due to the larger density of monomers in between the two cores. To improve the similarity between star-like microgels and star polymers, we thus considered stars with a partial core coverage, finding remarkable agreement between the two also in terms of quantitative effective potential. 

These findings align with expectations that microgels synthesized by a precipitation polymerization process around a tiny central core do not have the ability to fill the core coverage, thus remaining partially naked. Given that the partial core coverage was not previously considered in the literature, we also provided evidence that it is a minor difference in the topology of the particles, not affecting the structural properties of the stars and the vast results already established in the literature, including the logarithmic dependence of the interparticle potential upon core-core distance. It will be interesting in the future to compare the present results with alternative types of microgels with star-like characteristics, resulting from a more complex synthetic method developed in the group of Kanaoka~\cite{ida2020thermoresponsive}, that have not yet been described in terms of theoretical models.

To conclude, star-like microgels are viable alternatives to star polymers, with very similar single-particle properties and two-body effective interactions, that will be extremely interesting to investigate in the future. Indeed, the coupling of their ultrasoft character with their enhanced softness, as compared to standard microgels, may provide interesting results for phase behavior and rheological response in dense suspensions.

\section*{Acknowledgements}
The authors thank Simona Sennato, Jacopo Vialetto, Marco Laurati and Leah Rank for useful discussions. TP, EB and EZ acknowledge financial support from Progetto Co-MGELS funded by the European Union - NextGeneration EU under the National Recovery and Resilience Plan (PNRR) Mission 4 “Istruzione e Ricerca” - Component C2 - Investment 1.1 - "Fondo PRIN", Project code PRIN2022PAYLXW Sector PE11, CUP B53D23008890006. FB and EZ also acknowledge support from ERC POC project MICROSENS (grant agreement no.101157420). EZ also acknowledges funding from ICSC – Centro Nazionale di Ricerca in High Performance Computing, Big Data and Quantum Computing, funded by European Union – NextGenerationEU - PNRR, Missione 4 Componente 2 Investimento 1.4. We gratefully acknowledge the CINECA award under the ISCRA initiative, for the availability of high-performance computing resources and support.


\bibliography{star-polymers}

\clearpage
\newpage
\onecolumngrid

\begin{center}
\large
\textbf{Supplementary Information for: Star-like microgels vs star polymers: similarities and differences}

\normalsize
\bigskip
Tommaso Papetti\textsuperscript{ 1, 2}, Elisa Ballin\textsuperscript{ 1, 2}, Francesco Brasili\textsuperscript{ 2, 1}, Emanuela Zaccarelli\textsuperscript{ 2, 1}\\
\medskip
\small
\textit{%
\textsuperscript{1}Department of Physics, Sapienza University of Rome, Piazzale Aldo Moro 2, 00185, Roma, Italy\\
\textsuperscript{2}CNR Institute of Complex Systems, Uos Sapienza, Piazzale Aldo Moro 2, 00185, Roma, Italy\\
}
    
\end{center}

\normalsize\bigskip
\renewcommand{\thefigure}{S\arabic{figure}}
\setcounter{figure}{0}

\section{Representative Snapshots}

\begin{figure}[h!]
    \centering
    \includegraphics[width=1\linewidth]{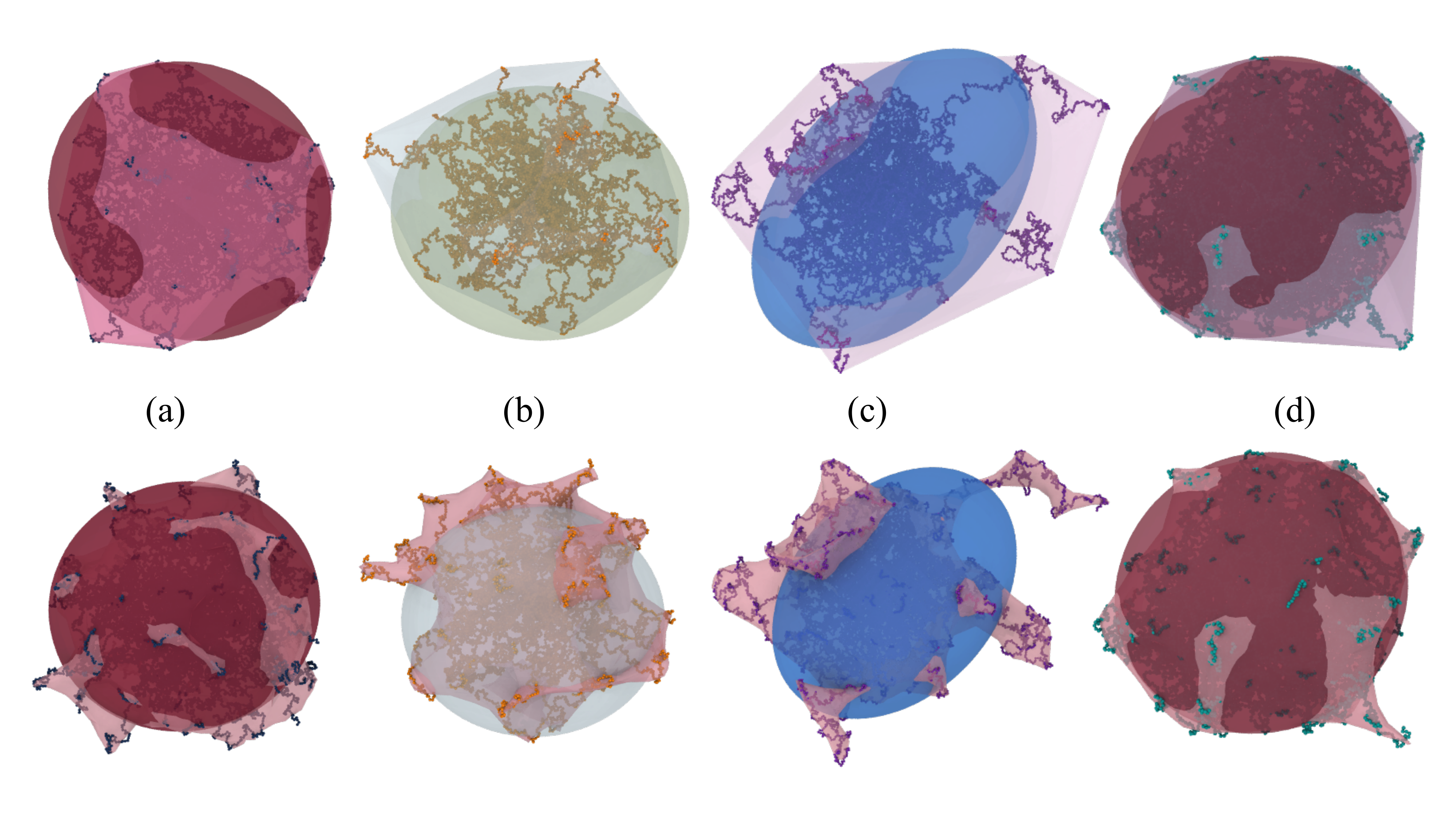}
    \caption{Snapshots for various particles analyzed in the paper. The top row represents the particles together with their convex hull mesh and relative effective ellipsoid, the bottom row is analogous but referring to the surface mesh method. (a): star polymer with $f = 80$, $N_f = 200$; (b): partially covered star polymer with $f = 46$, $N_f = 200$, $\gamma = 0.57$; (c): star-like microgel with $c = 1 \%$ and (d) standard microgel with $c = 1 \%$.} 
    \label{fig:snap_methods}
\end{figure}

In Fig.~\ref{fig:snap_methods} we present representative snapshots for the different types of soft particles analyzed in the main text, to illustrate how we estimate their volume. In particular, each snapshot compares the volumes obtained with the convex hull approximation (top row) and the surface mesh approach (bottom row), along with the corresponding effective ellipsoids.

The Figure shows that the ellipsoids obtained from the surface mesh method are systematically smaller than those derived from the convex hull approach. This difference is particularly evident for the star-like microgel (c), highlighting that the convex hull method is especially sensitive to conformations featuring occasional long outer chains. In addition, the star-like microgel appears to be most asymmetric among the studied particles, again due to the presence of the long outer chains.
Given that the surface mesh seems to give a more accurate estimate of the particle volume, we then mostly use this method in the main text.

\section{Experimental measure of $R_g$}

\begin{figure}[h!]
    \centering
    \includegraphics[width=0.6\linewidth]{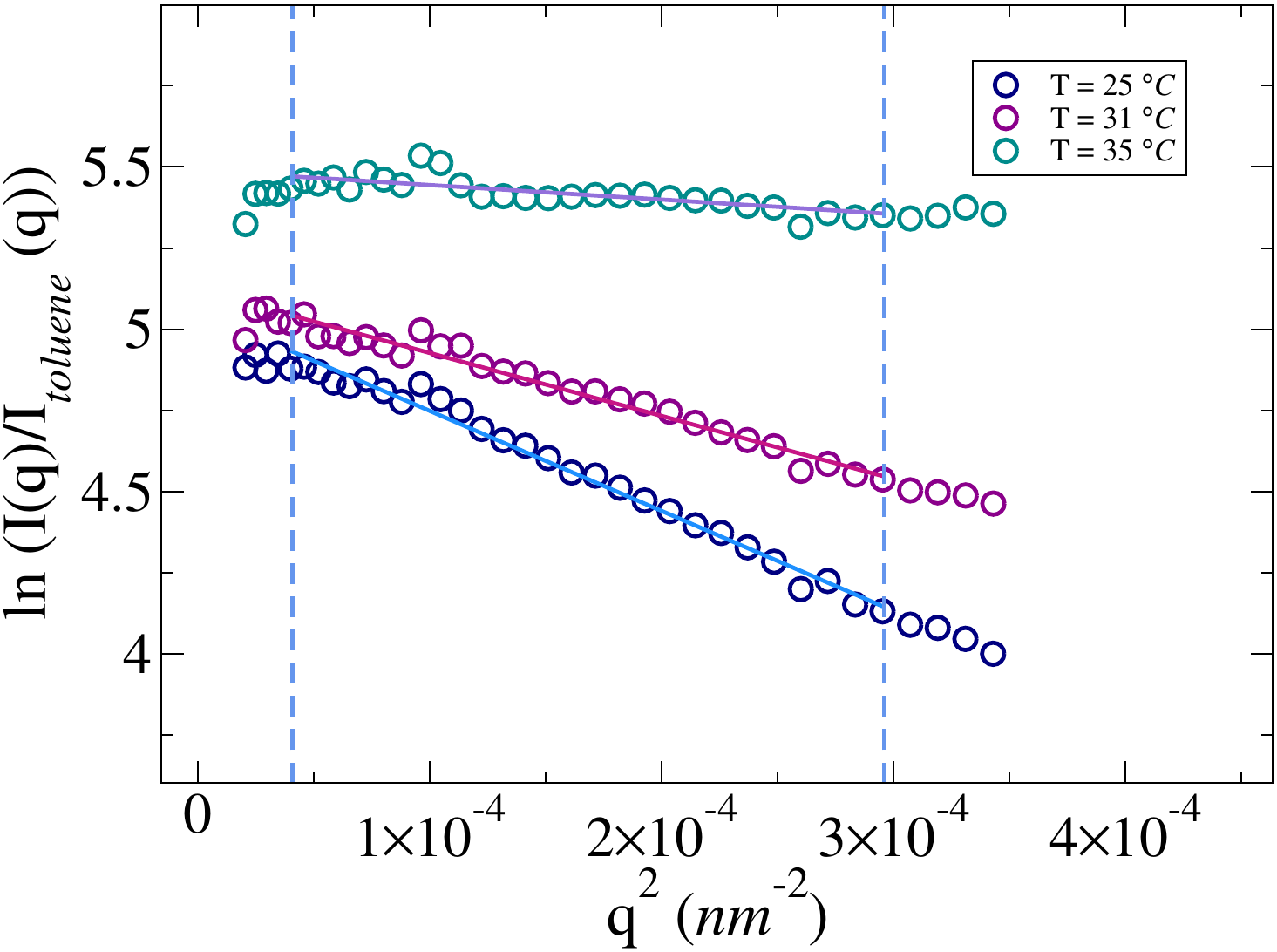}
    \caption{Selected Guinier fits for the estimation of $R_g$ from SLS data at three studied temperatures for PNIPAM-EGDMA microgels with c=1\%. Vertical dashed lines are for visual reference to the fit interval.}
    \label{fig:guinier_comparison}
\end{figure}

As explained in the main text, the radius of gyration $R_g$ was measured from static light-scattering (SLS) by fitting the angular-dependent scattered intensity with the Guinier form:
\begin{equation}
\label{eq:Sguinier}
    I(q)=I(0)\exp\bigg[-\frac{(qR_g)^2}{3}\bigg]\, .
\end{equation}
The measured intensity $I(q)$ was normalized by the scattered intensity of Toluene, $I_{\mathrm{toluene}}(q)$, and analyzed in the Guinier representation, i.e., by plotting $\ln\!\left[I(q)/I_{\mathrm{toluene}}(q)\right]$ as a function of $q^2$. Representative fits of the SLS data at different temperatures are reported in Fig.~\ref{fig:guinier_comparison}.

\section{Difference in the choice of variable (COM vs core) for effective interactions of star polymers}

\begin{figure}[h!]
  \centering
  \includegraphics[width=0.6\linewidth]{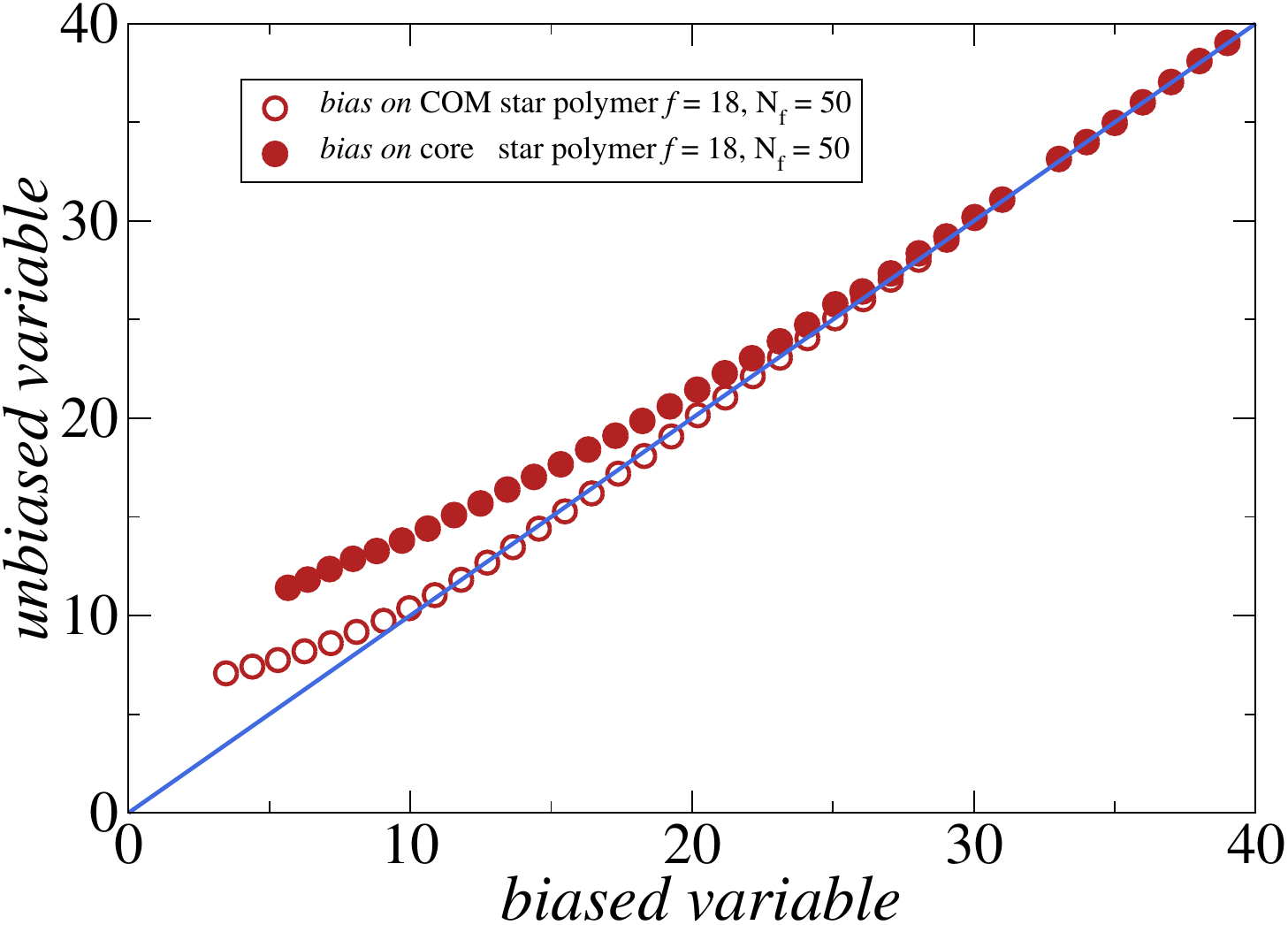}
  \caption{Behavior of the mean separation between natural variables for the umbrella sampling windows of the star polymer with $f = 18$, $N_f = 50$. The $x$-axes represents the biased variable, i.e., the natural variable chosen for the potential - either $r_{COM}$ or $r_{core}$ -  while the $y$-axes shows the mean distance of the other respective variable averaged over the simulation for each of the windows. The blue line is the bisector.}
  \label{fig:biasedVSunbiased}
\end{figure}

The choice of variable has a significant impact on the effective two-body interaction between star polymers. Namely, the potentials expressed as a function of the distance between the cores or between the centers of mass have different functional forms. This is demonstrated by our umbrella sampling simulations, which can be perfomed biasing one or the other variable. As shown in Fig.~\ref{fig:biasedVSunbiased}, 
where we plot, for each umbrella window, the mean value of the unbiased distance versus the biased one, the two constraints used in the context of umbrella sampling technique select qualitatively different regions of configurational space over a substantial range of separations.

This asymmetry is twofold: the separation from the bisector is smaller for the COMs bias and takes place at closer separations, while the bias-on-cores line separates widely from the bisector and early on, beginning even at $r \sim 2.4 R_g$, indicating that the conditional distributions $P(r_{\mathrm{COM}}\!\mid r_{\mathrm{core}})$ and $P(r_{\mathrm{core}}\!\mid r_{\mathrm{COM}})$ are broad and, crucially, not equivalent in the range where the two macromolecules start to mutually deform and interpenetrate. As a consequence, a simple functional mapping between the two distances such as $V(r_{\mathrm{COM}})=V[r_{\mathrm{core}}(r_{\mathrm{COM}})]$ cannot be used to reconcile the difference. 

\begin{figure}[h!]
  \centering
  \includegraphics[width=0.7\linewidth]{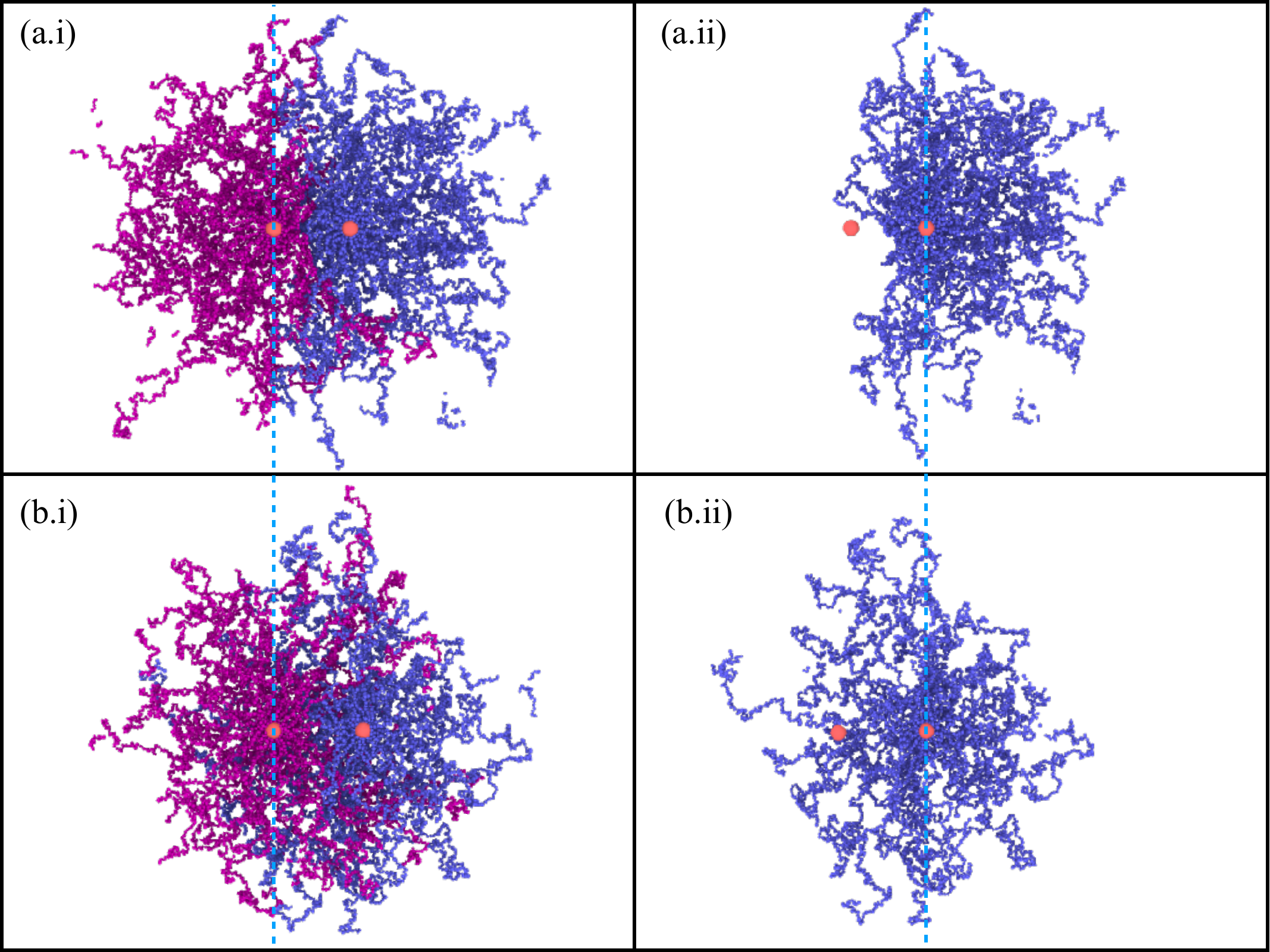}
  \caption{Equilibrium snapshots of two star polymers with $f = 80$, $N_f =200$ taken during the umbrella sampling window at a separation value $20$ in each of the two variables: (a) $r_{core}$ and (b) $r_{COM}$. The second column (ii) shows the detail of only a single particle in the same configuration, hiding the other particle except its core. 
  Note that the cores in the second row are found at a separation larger than $20 \sigma$, since the bias is on $r_{COM}$. The vertical dashed lines are guides to the eye.}
  \label{fig:coreVScom_snapshots}
\end{figure}
The microscopic origin of these deviations is illustrated by the snapshots in Fig.~\ref{fig:coreVScom_snapshots}. Indeed,
constraining $r_{\mathrm{core}}$ directly controls the separation of the anchoring regions of the chains, the cores, where crowding is strongest, while leaving the mass distribution of the arms free to reorganize around that constraint, so that they move away from the central zone. In contrast, constraining $r_{\mathrm{COM}}$ can be satisfied through internal rearrangements of the chains: the stars redistribute their mass, so that the centers of mass approach each other without requiring an equally small core–core distance. This results in a more crowded monomer-rich central region when the bias is applied to the COM, and consequently $V_{\mathrm{eff}}$ (see Fig.1 of the main text) is more repulsive over a substantial range of separations. Clearly, the two potentials eventually cross: the cores are physical points and the potential diverges upon core contact, whereas the centers of mass are fictitious points, and their overlap is, in principle, allowed.

\section{Swelling curves}
\begin{figure}[h!]
    \centering
    \includegraphics[width=0.48\linewidth]{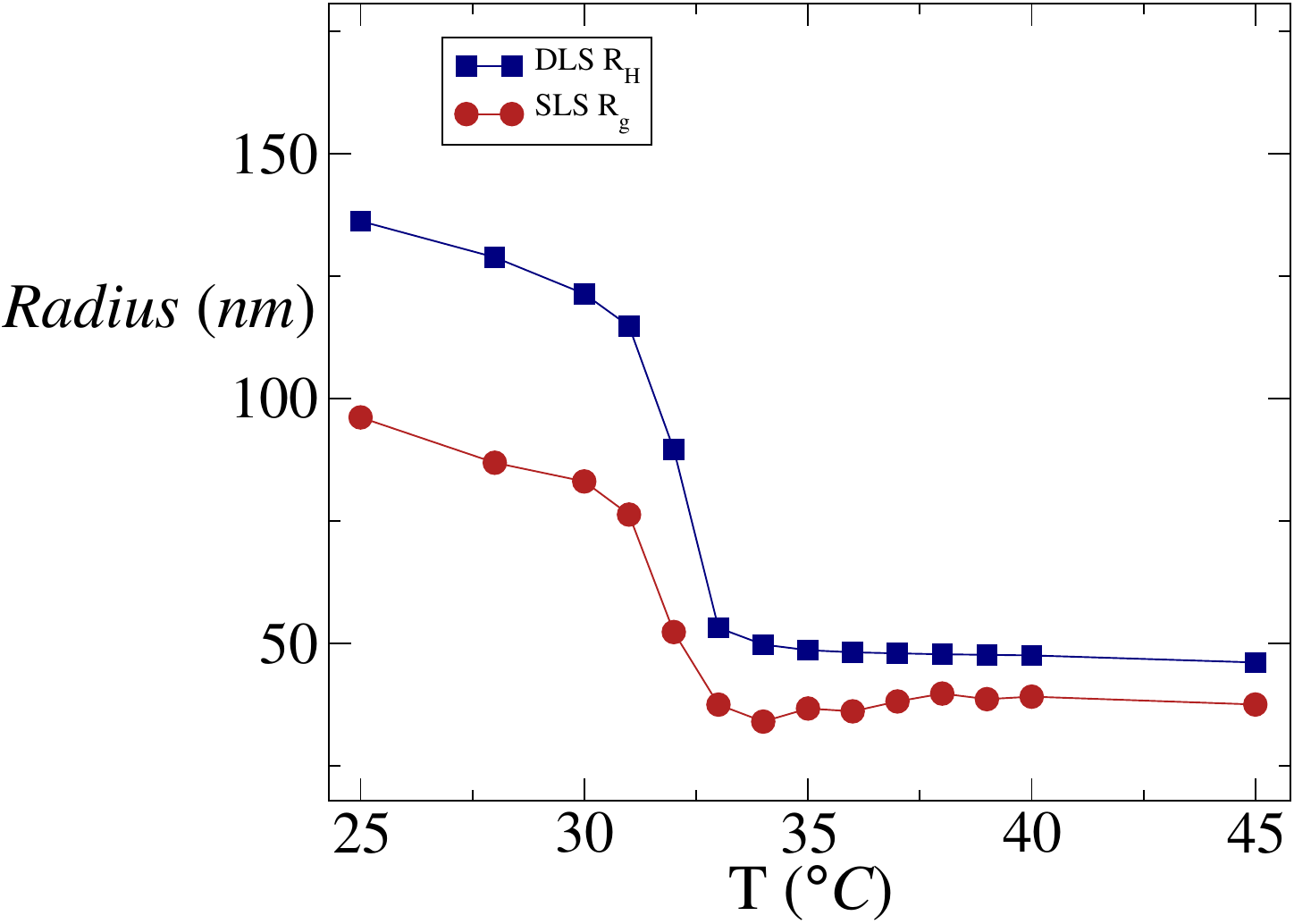}
    \caption{Swelling curves for the SLS ($R_g$) and DLS ($R_H$) data of the star microgel with $c=1\%$ concentration of crosslinkers.}
    \label{fig:single/swelling_experimental}
\end{figure}

\begin{figure}[h!]
    \centering
    \includegraphics[width=0.48\linewidth]{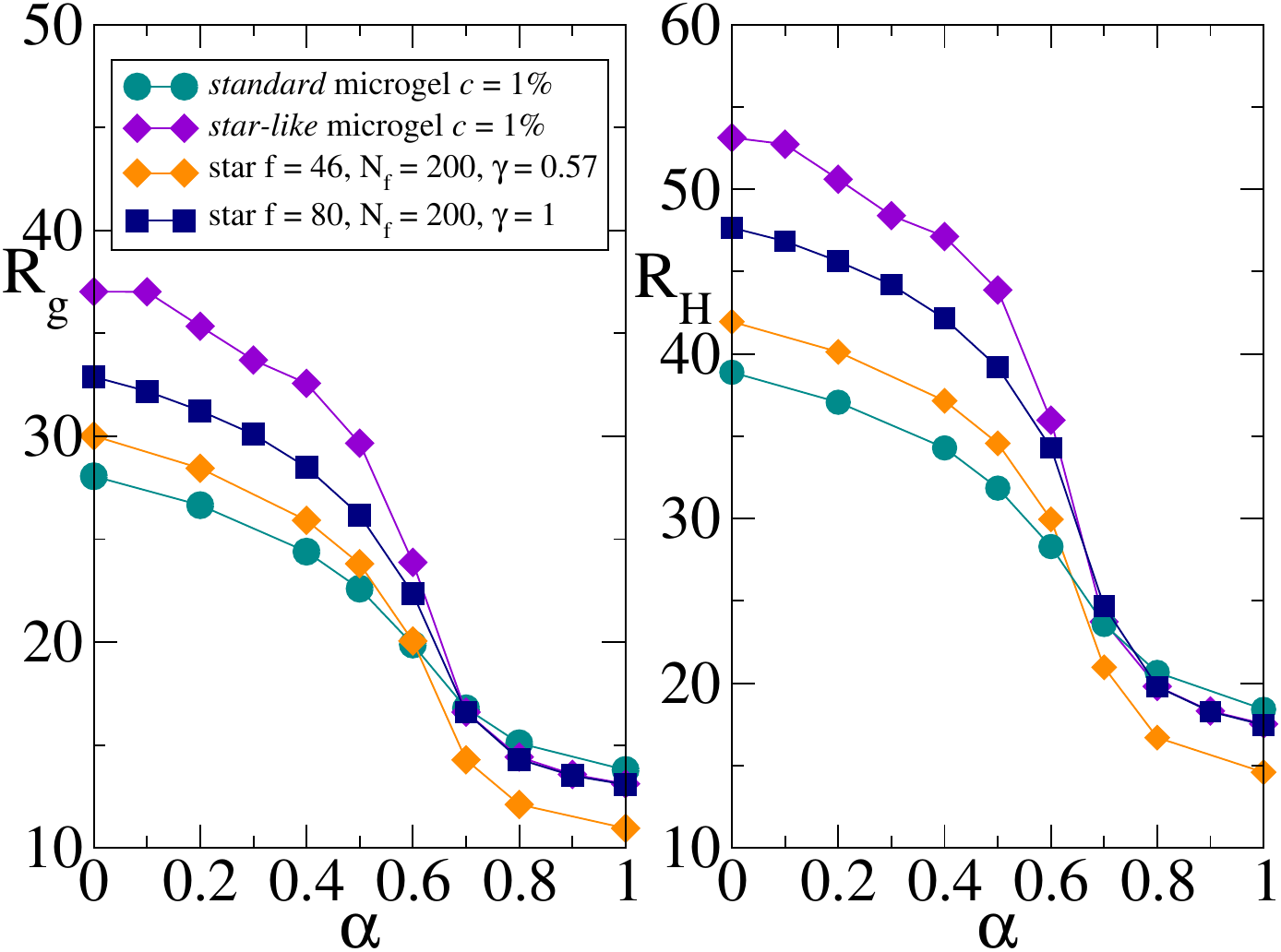}
    \caption{Swelling curves for the different simulated particles. Here, $R_H$ is calculated within the surface mesh method.}
    \label{fig:single/swelling}
\end{figure}

For completeness we also report the swelling curves for $R_H$
from DLS data and $R_g$ from SLS data in Fig.~\ref{fig:single/swelling_experimental} and compare them with the numerical swelling curves of Fig.\ref{fig:single/swelling}. The latter also reports data for the additional types of particles studied in the simulations.

Notably, for star-like microgels, in both experiments and simulations, the reduction of $R_H$ is more pronounced than that of $R_g$, indicating that the hydrodynamic size is particularly sensitive to the rearrangement of the outer, more dilute corona.

\newpage
\section{Numerical results from the convex hull approach}

Finally, we report the convex-hull-based estimates of the ratio $R_g/R_H$  and of the bulk modulus, complementing the surface-mesh results discussed in the main text.

\begin{figure}[h!]
  \centering
  \includegraphics[width=0.48\linewidth]{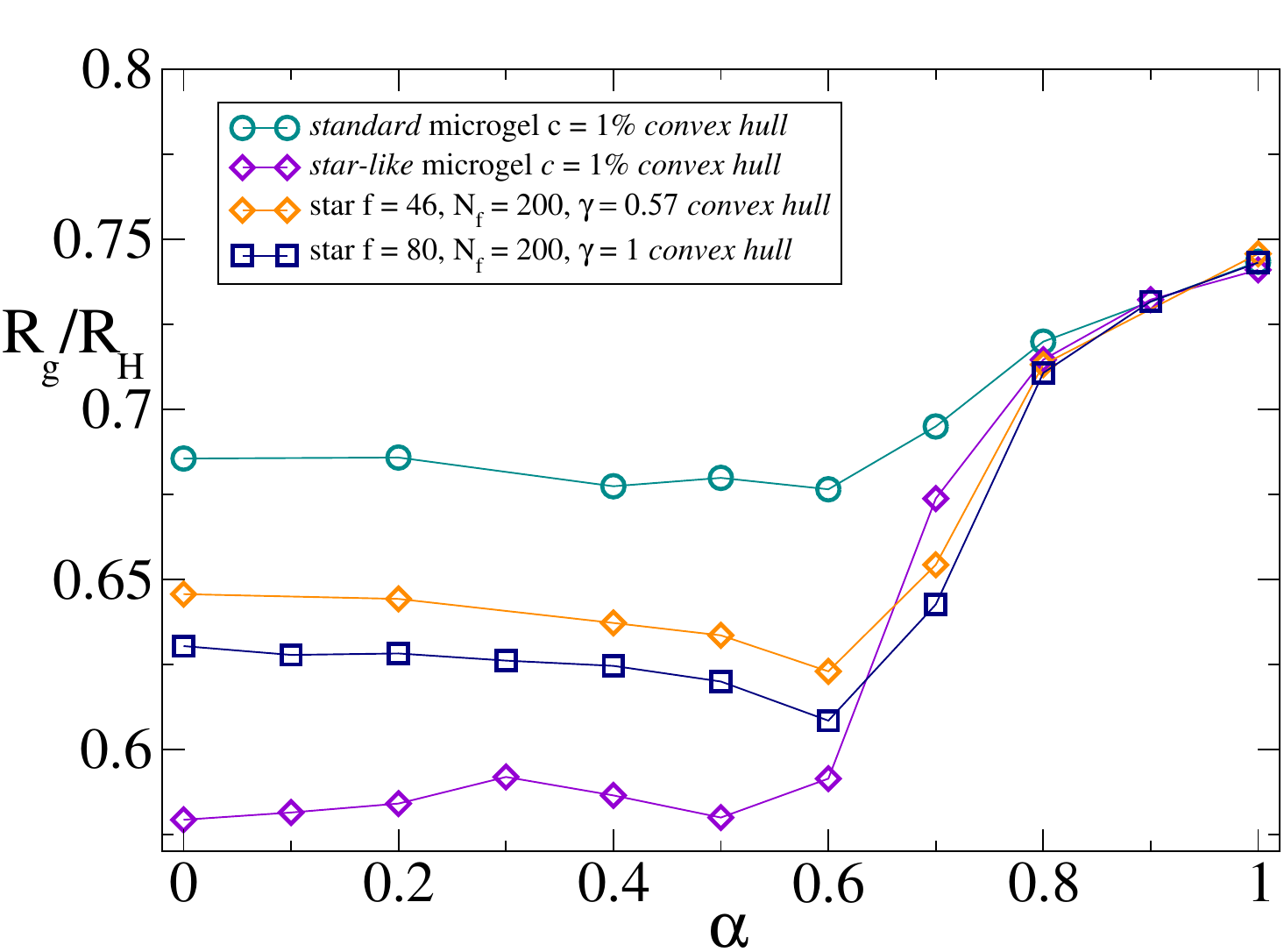}
  \caption{Ratio $R_g/R_H$ for the \emph{star-like} microgel with $c=1\%$, for the standard microgel also with $c=1\%$, for the fully covered star ($f = 80$, $N = 200$) and for the partially covered star with $f = 46$ and $N = 200$ ($\gamma \sim 0.57$). $R_H$ is calculated with the convex hull method.}
  \label{fig:Rg_Rh_ch}
\end{figure}

The ratio $R_g/R_H$ obtained by computing $R_H$ from the convex hull is shown in Fig.~\ref{fig:Rg_Rh_ch}. While the overall evolution with $\alpha$ is consistent with a progressive collapse across the VPT region, the convex hull amplifies differences between architectures in the good-solvent regime: the star-like microgel displays the strongest deviation because even a small number of extended chains can substantially enlarge the convex envelope and thus increase the inferred $R_H$.

\begin{figure}[h!]
  \centering
  \includegraphics[width=0.48\linewidth]{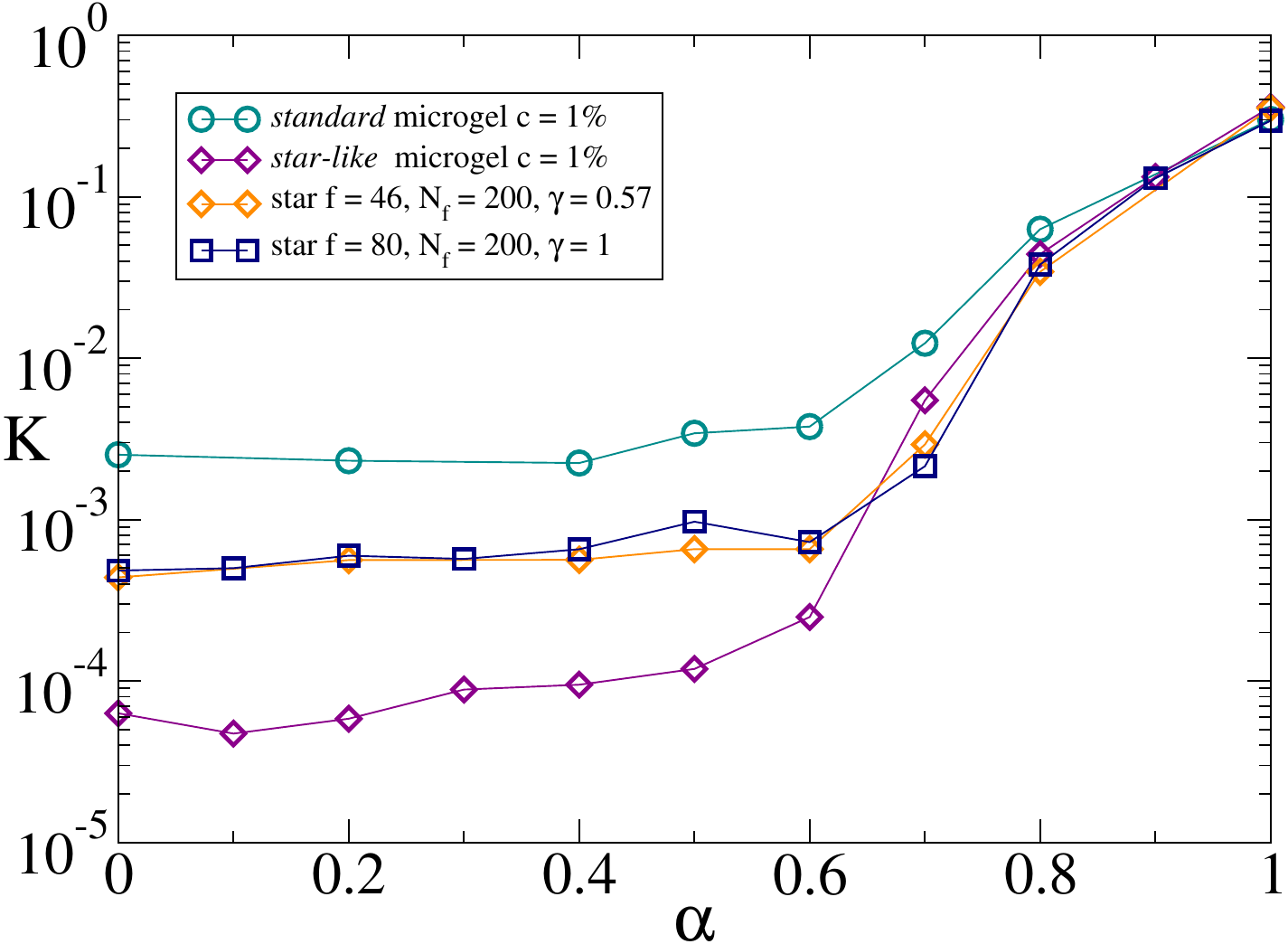}
  \caption{Bulk modulus $K$ in units of $k_BT/\sigma^3$ as a function of effective temperature $\alpha$, obtained using the convex hull method to estimate the volume. Note that this 
method yields systematically smaller values of $K$ for the \emph{star-like} microgel, due to its sensitivity to occasional long, sparse chains.}
  \label{fig:bulk_modulus_ch}
\end{figure}

Figure~\ref{fig:bulk_modulus_ch} reports the bulk modulus $K$ as a function of $\alpha$ when the particle volume is estimated from the convex hull. The transition region is still clearly identified by the sharp increase of $K$ upon collapse. However, the star-like microgel yields systematically smaller $K$ values with respect to the other systems. This is consistent with the convex hull overestimating the instantaneous volume (and its fluctuations) whenever sparse protrusions are present, which in turn biases the inferred compressibility. These convex-hull-based results therefore provide a useful cross-check, while also highlighting why the surface mesh method is adopted as the reference characterization in the main text.

\end{document}